\documentclass[12pt,preprint,twoside]{aastex}

\shortauthors{Collins et al.}
\shorttitle{Narrow-Line Region of Markarian 3}

\begin{document}

\title{Physical Conditions in the Narrow-Line Region of Markarian 3.
II. Photoionization Modeling Results\altaffilmark{1}}

\author{N.R. Collins\altaffilmark{2},
S.B. Kraemer\altaffilmark{2}, 
D.M. Crenshaw\altaffilmark{3},
F.C. Bruhweiler\altaffilmark{2} \& 
M. M\'{e}lendez\altaffilmark{4}}

\altaffiltext{1}{Based on observations made with the NASA/ESA {\it
Hubble Space Telescope}, obtained at the Space Telescope Science
Institute, which is operated by the Association of Universities for
 Research in Astronomy, Inc., under NASA contract NAS 5-26555}

\altaffiltext{2}{Institute for Astrophysics and Computational
Sciences, Department of Physics, The Catholic University of America;  
and Astrophysics Science Division, Code 667,
NASA Goddard Space Flight Center, Greenbelt, MD  20771; 
nicholas.collins@nasa.gov, steven.b.kraemer@nasa.gov, 
frederick.c.bruhweiler@nasa.gov}

\altaffiltext{3}{Department of Physics and Astronomy, Georgia State
University, Atlanta, GA 30303}

\altaffiltext{4}{NASA Postdoctoral Program Fellow, 
Goddard Space Flight Center, Code 662,
Greenbelt, MD 20771; marcio@milkyway.gsfc.nasa.gov}

\begin{abstract}

We have examined the physical conditions in the narrow-line region
(NLR) of the Seyfert 2 galaxy Markarian~3, using long-slit spectra
obtained with the {\it Hubble Space Telescope}/Space Telescope Imaging
Spectrograph and photoionization models. We find three components of
photoionized gas in the NLR. Two of these components, characterized by
emission lines such as [Ne~V] $\lambda$3426 and [O~III] $\lambda$5007,
lie within the envelope of the bi-conical region described in our
previous kinematic study. A component of lower ionization gas, in
which lines such as [O~II] $\lambda$3727 arise, is
found to lie outside the bi-cone. Each of these components is
irradiated by a power-law continuum which is attenuated by
intervening gas, presumably closer to the central source. The
radiation incident upon the low ionization gas, external to the
bi-cone, is much more heavily absorbed. These absorbers are similar to
the intrinsic UV and X-ray absorbers detected in many Seyfert 1
galaxies, which suggests that the collimation of the ionizing
radiation occurs in a circumnuclear wind, rather than a thick,
molecular torus.  We estimate the mass for the observed NLR emitting 
gas to be $2~\times~10^{6}~M_{\odot}$.  It is likely that Markarian~3 
acquired this gas through an on-going interaction with the  
spiral galaxy UGC~3422.  

\end{abstract}

\keywords{galaxies: individual (Markarian 3) -- galaxies: Seyfert --
line: formation}
~~~~~

\section{Introduction}\label{intro}

Markarian~3, which is among the brightest Seyfert 2 galaxies, is
classified as a Hubble-type SB0 galaxy \citep{ada77}. Its systemic
velocity is 4050~km~s$^{-1}$ ($z$~=~0.0135) based on \ion{H}{1}~21~cm
emission \citep{tif88}, which yields a distance of 53~Mpc for
$H_{0}$=75 km s$^{-1}$ Mpc$^{-1}$.  At this distance, 1$\arcsec$
corresponds to 257~pc. Spectropolarimetry of Markarian~3 revealed
broad permitted lines and non-stellar continuum emission
\citep{sch85}. These observations are among those cited as evidence for
the ``unified'' model for Seyfert galaxies \citep{ant93}, which posits
that the central engine and broad emission-line region (BLR) are
surrounded by a dense molecular torus. In this model, when we observe a
Seyfert~2 galaxy our line-of-sight is occluded by the torus and the
central region is hidden.  The Seyfert~2 active nucleus and BLR 
may sometimes be detected via radiation
scattered into our line of sight by free electrons in the narrow
emission-line region (NLR). In principle, one can probe into the
nature of the hidden active galactic nucleus via a detailed study of
the NLR gas.

In our earlier paper \citep[][; hereafter Paper I]{col05}, we used {\it
Hubble Space Telescope}/Space Telescope Imaging Spectrograph ({\it
HST}/STIS) longslit low-resolution spectroscopy from 1150~\AA~ to
10,300~\AA~ to study the physical conditions in the narrow-line NLR of
Markarian~3 (we refer the reader to Paper I for a more extensive
discussion of earlier studies of Markarian~3). Here, we summarize the
main results of Paper I.  We found that the extinction within
Markarian~3 along the line-of-sight to the NLR is best characterized
by a Large Magellanic Cloud (LMC) type extinction curve \citep{koo81}.
There is an extinction gradient increasing from West to East along the
STIS slit (at position angle 71$\degr$ measured East from North) in
both line and continuum emission.  We interpreted this gradient as
evidence that the host galaxy disk is tilted towards the observer in
the east.  From emission-line diagnostics we argued that the NLR is
photoionized by the hidden active galactic nucleus (AGN) continuum and
that the density of the NLR gas decreases with increasing distance
from the center.  We modeled the observed continuum as a combination
of reddened host galaxy light from an old stellar population, reddened
H$^{+}$ and He$^{+2}$ recombination continua, and scattered light from
the central engine with spectral index $\alpha$=1
($L_{\nu}\propto\nu^{-\alpha}$).  The host galaxy to scattered-light
ratio was estimated to be 3:1 at 8125~\AA~ in a 0$\farcs$1 $\times$
1$\farcs$8 aperture.  We fitted the intrinsic ionizing continuum with
a two-component power-law of the form $L_{\nu} \propto \nu^{-\alpha}$,
where $\alpha$=2 for 13.6~eV~$<$~$E$~$<$~0.2~keV and $\alpha$=1 for
0.2~keV~$<$~$E$~$<$~50~keV). Based on this analysis, we estimated that the amount
of intrinsic non-ionizing UV continuum scattered into our
line-of-sight is 0.04\%.

\citet{kra86} analyzed a combination of {\it International Ultraviolet
Explorer} and ground-based spectroscopic observations of the
Markarian~3 NLR, from ultraviolet to infrared wavelengths. Based on
this analysis, they found evidence for three components of
emission-line gas. A single component would be insufficient to
reproduce all the observed emission lines 
which arise from ionic species with ionization potentials up to 100~eV.
Two of their model components included internal dust.  The third component, 
which was  the highest in ionization and density and closest to the
central source, was assumed to be dust-free. Dust emission signatures
are evident in the infrared continuum at $\lambda~\sim~$10$\mu$m
\citep{neu76,rie78,wee05}. \citet{kra86} argued that these arose
from hot dust within the two lower-density components. Since dust
suppresses resonance line emission with large optical depths 
and the presence of dust requires depletion of C and Mg from 
the gas phase, 
they
included the dust-free component to match the observed
Ly$\alpha$~$\lambda$1216, \ion{C}{4}~$\lambda\lambda$1548,1551 and
\ion{Mg}{2}~$\lambda\lambda$2796,2803 emission. In this study, we
expand on these results through a photoionization modeling analysis of
the STIS long-slit spectra presented in Paper I.  
This analysis has enabled us to determine the physical conditions
as a function of radial distance, constrain the energetics of the
central source, and develop new insight into the relationship between
the structure of the NLR and the conditions in the unresolved
circumnuclear region.

\section{Observations and Analysis}\label{observations}

The details of the observations and data reduction are described in
Paper I. To summarize, \emph{HST}/STIS longslit spectra were obtained
on 2000 August 22 in all four low-resolution modes (G140L, G230L,
G430L, and G750L) providing wavelength coverage from 1150~\AA~ to
10,300~\AA.  The 52$\arcsec\times$0$\farcs$1 slit was used for all
spectral modes.  The slit was oriented to position-angle 71$\degr$
East of North to match the inner portion of the elongated
reverse-``S'' shaped feature in the brightest part of the central NLR
(see Figure 1 in Paper I).  We made seven discrete measurements of
each emission line in the spatial direction.  
Individual spectra were
extracted by summing array rows in selected bins in the spatial
direction.  The bin sizes were selected to extract spectral
information for individual clouds and to achieve a signal-to-noise
ratio greater than 5 for the
\ion{He}{2}~$\lambda$1640/$\lambda$4686 measurement: as discussed
below, this ratio was used to determine the extinction along the
line-of-sight to the NLR.  The bin sizes range from 0.2-0.3
arcseconds, or 52-77 pc, using the adopted distance scale described in
\S\ref{intro}.  In some measurement bins there are two kinematic
components and/or two different lines are blended together.  For those
cases we used parameters (kinematic component velocity and velocity
dispersion) derived from the $[$\ion{O}{3}$]$~$\lambda$5007 line in
the same bin as a template to separate the individual components.

We note here an error in the identification of the central NLR bin, 
that bin containing the AGN.  In Paper~I Figure~6 the AGN is in the 
bin to the east of the bin indicated by the arrow (cf. Paper~I Figure~1 
which shows the correct identification).  This error does not affect 
any of our conclusions in that paper.  However, for the corrected 
plots vs. position (in arcseconds) along the slit  all points would 
move towards the west.   The bin labels in Paper~I Tables~2 and 3 would be 
similarly corrected to the west as they are in \S\ref{modelfits} of 
this paper.  

We corrected the observed flux measurements for both Galactic
extinction along the line-of-sight to Markarian~3, and intrinsic
extinction caused by dust within Markarian~3.  The Galactic
color-excess of $E(B-V)$~=~0.19 was obtained from the dust map of
\citet{sch98}. We used the \citet{sav79} extinction curve to correct
for Galactic extinction.  We used the
\ion{He}{2}~$\lambda$1640/$\lambda$4686 line ratio \citep{sea78} to
determine the extinction within Markarian~3 along the line-of-sight.
This ratio has a longer wavelength baseline than the Balmer line ratio
(H$\alpha$~$\lambda$6563/H$\beta$~$\lambda$4861) and therefore gives a
better estimate of the reddening.  It is also relatively insensitive
to collisional excitation. As noted in Paper I, we found that the
LMC curve of \citet{koo81} gave the best agreement for the
\ion{He}{2}~$\lambda$1640/$\lambda$4686 and Balmer ratios with their
theoretical values. Therefore, we used this curve and a screen
geometry to de-redden the observed fluxes in this paper.  
The extinction corrected line fluxes relative to 
H$\beta$~$\lambda$4861 are listed in Table~3 of Paper I.

\section{Photoionization Models}\label{photomodels}

\subsection{Preliminary Model Input Parameters}\label{modelinput}

In this study we used version 95.06 of the photoionization modeling
code CLOUDY \citep{fer98}. As per convention, the photoionization
models are parameterized in terms of the ionization parameter, $U$:
\begin{equation}
U = \frac{Q(H)}{4 \pi r^2 n_{H} c}
\end{equation} 
where $Q(H)$ is the ionizing photon luminosity of the AGN in
photons~s$^{-1}$, $r$ is the radial distance of the emitting cloud
from the ionizing radiation source (the AGN in this case), $n_{H}$ is
the hydrogen number density in cm$^{-3}$ and $c$ is the speed of
light.  The photon luminosity and the spectral energy distribution are
related by
\begin{equation} 
Q(H) = \int_{\frac{13.6eV}{h}}^{\infty} \frac{L_{\nu}}{h \nu} d\nu
\end{equation}
In Paper I, we parameterized the intrinsic ionizing continuum as
follows: $L_{\nu}$~$\propto$~$\nu^{-\alpha}$, where $\alpha$~=~1 for
$h\nu$~$<$~ 13.6~eV, $\alpha$~=~2 for 13.6~eV~$<$ $h\nu$~$<$~200~eV,
and $\alpha$~=~1 for $h\nu$~$>$~200~eV, and $h$ is Planck's constant.
The high energy continuum slope is from the \citet{tur97} estimate
(based on $ASCA$ data) of the unabsorbed X-ray spectrum, with
normalization such that the unabsorbed 2-10~keV integrated luminosity
is 10$^{44}$~ergs~s$^{-1}$.  This is consistent with the intrinsic
luminosity derived by \citet{mat00} with $BeppoSAX$ data and the
prediction of \citet{mel08} extrapolated from a $Spitzer$
$[$\ion{O}{4}$]$~$\lambda$25.89~$\mu$m measurement.  We truncated the
continuum on the high energy end at 10$^{5}$~eV. Based on this
parameterization, the intrinsic AGN continuum is extremely luminous,
with $Q(H)$~$\sim$~2$\times$10$^{55}$~photons~s$^{-1}$.

We assume roughly solar chemical composition \citep[e.g.][]{gre89} for
all model components and absorbing screens 
(described in \S\ref{modelcomponents}).  The abundances
relative to H by number are:  He~=~0.1,
C~=~3.4~$\times$~10$^{-4}$,
N~=~1.2~$\times$~10$^{-4}$,
O~=~6.8~$\times$~10$^{-4}$,
Ne~=~1.1~$\times$~10$^{-4}$,
Mg~=~3.3~$\times$~10$^{-5}$,
Si~=~3.1~$\times$~10$^{-5}$,
S~=~1.5~$\times$~10$^{-5}$,
Ar~=~4.0~$\times$~10$^{-6}$ and
Fe~=~4.0~$\times$~10$^{-5}$.

For model components that included dust (see \S\ref{modelcomponents}) we used
elemental depletions onto graphite and silicate grains that are half
those of the standard Galactic interstellar medium (ISM) depletions.
The Galactic ISM depletion fractions are 0.65 for C and 0.50 for O
\citep{sno96}; 0.90 for Al, 0.44 for S, 0.90 for Ca and 0.90 for Ni
\citep{sea83}.  \citet{sno96} list the ISM depletion for Fe, Mg and Si
as 100\%.  We used a depletion of 95\% for these elements. 
We do not account for nitrogen depletion, since nitrogen is
deposited onto grains in ice mantles that would dissociate in the NLR.
The gas phase abundances in the dusty model component are the nominal
ISM values listed above multiplied by 0.5 $\times$ (1 -
$f_{depletion}$) where $f_{depletion}$ represents the Galactic ISM
depletion factor for the element of interest.

Our primary constraint on the models is that they reproduce the
observed emission line fluxes.  We required that the predicted fluxes
of at least three-quarters of the twenty selected bright emission
lines, each scaled to the predicted H$\beta$~$\lambda$4861 flux, match
the data within a factor two.  
We also required that at least half of
those lines within that gross limit match the data within $\pm$30\%.
It is possible to construct models in which the emission line fluxes
match but the emitting clouds\footnote{We refer to the regions over
which we have summed the flux, then separated into individual
kinematic components, as ``clouds''.}  may be unrealistically large.
We implemented additional constraints to ensure that the models fit
within our geometrical picture of the NLR and within the observational
parameters of the \emph{HST}/STIS spectra.

\subsection{Initial View of the NLR Structure}\label{nlrstructure1}

We initially assumed an NLR geometry based on the results of
\citet{rui01}. They studied the kinematics of the Markarian~3 NLR
using $[$\ion{O}{3}$]$~$\lambda$5007 measurements from a STIS/CCD
slitless spectrum and from the same long-slit data that was presented
in Paper I.  Extracted spectra from several locations along the
long-slit show evidence of two kinematic components: one is redshifted
and the other is blueshifted in the rest-frame of Markarian~3.  These
components are interpreted as line-of-sight observations of radially
outflowing gas on opposite sides of a NLR bi-cone.  Their best-fit
kinematic model assumes an NLR with an angular extent of 2$\arcsec$
and inner and outer opening half-angles of 15$\degr$ and 25$\degr$,
respectively.  The bi-cone is tilted toward the observer in the East
and away from the observer in the West by 5$\degr$.  The position
angle of the the bi-cone is 70$\degr$ East of North \citep{sch00}.
The host galaxy disk major-axis position angle is 28$\degr$ and the
inclination of the disk is 33$\degr$ \citep{sch00}.  We converted the
projected angular separation in arcseconds between the central engine
and a measured kinematic component in the NLR into a radial distance
using the bi-cone model opening angles and the 257~pc~arcsecond$^{-1}$
scale factor from our estimated distance to Markarian~3.  Redshifted
kinematic components lie on the far side of the bi-cone and
blueshifted components lie on the near side.  For the central bin
(hereafter angular position 0$\farcs$0) we assumed radial
distances from the AGN for the redshifted and blueshifted components
corresponding to one-quarter of the total 0$\farcs$3 bin width.

We required that the model components fit 
within the intersection of the STIS slit, 
our measurement bins, and the bi-cone geometry.
These three constraints correspond to the 
dispersion direction on the plane of the sky, 
the cross-dispersion direction on the plane of the 
sky, and the direction perpendicular to the sky plane, 
respectively.  Each measurement bin may be 
considered as a small cell constrained, by the
the projected slit width (26~pc), 
the measurement bin height along the slit, 
and the inner and outer bi-cone walls.  

In this simple picture we envisioned a model 
component inside the measurement cell as a brick in which the illuminated face 
and the emitting face are the same.  The model illuminated 
face is parallel to the cell face defined by the 
slit width and the lines between the inner and outer 
bi-cone walls at the slit edges.  
The model emitting face area was computed as the ratio a:b of (a) the 
observed H$\beta$ luminosity (ergs~s$^{-1}$) of the 
measured kinematic component scaled by the fractional 
contribution of the  model component to (b) the 
predicted H$\beta$ flux (ergs~s$^{-1}$~cm$^{-2}$) for 
the model component.  This emitting area must be on the order of 
cell face area. 

We refer to the distance between the bi-cone walls as the ``bin depth''.
We calculate a ``model component depth'' by taking the ratio 
of the emitting area to the projected slit width in pc. 
It is unlikely that the bi-cone has a sharp cut-off. 
\citet{kra08} suggest that the NGC~4151 NLR does not have a sharp edge.  
Therefore, we allow the model component depth to be 
up to 1.5 times the ``bin depth'' at the center of the measurement bin.  
The height of the model component brick is given by the ratio 
of the model input parameters column density 
(in cm$^{-2}$) and volume number density (cm$^{-3}$).
We required that this height be less than  the measurement bin 
height along the STIS slit. 

As a check that all model components fit within 
a measurement bin we computed the volume filling 
factor of each component.  The volume filling factor 
is the ratio of the model component brick volume 
(emitting area multiplied by component height) 
to the measurement bin volume.  We required that 
the sum of the volume filling factors of the 
model components within a bin be less than unity.
We modified these geometrical constraints after analyzing the 
preliminary model results and reviewing the spectral 
image data (see \S\ref{modelinput}).

In Paper I,  we presented observational evidence that the
n$_H$ decreases with increasing distance from the
AGN.  For the modeling we did not assume a specific 
functional form for the radial dependence
of model cloud number density.  We simply required that a cloud must
have the same or lower number density than the neighboring cloud in
the direction of the AGN.

\subsection{Evidence for Emission-Line Gas Outside the Bi-cone}\label{spatialdiff}

As shown in Tables 2 and 3 in Paper I, each spectrum extracted along
the slit shows a mix of strong emission lines from high and low
ionization species, e.g., $[$\ion{Ne}{5}$]$~$\lambda$3426 and
$[$\ion{O}{2}$]$~$\lambda$3727 (the unresolved blend of the 3726.0
\AA~ and 3728.8 \AA~lines) emission, from the same kinematic
component.  This suggests that the observed emission along a line of
sight to the Markarian~3 NLR may come from a combination of co-located
clouds of different ionization states and optical thicknesses. Similar
multi-component models have been used to model the NLR emission in the 
Sy~1 NGC~4151 \citep{kra00} and the Sy~2 NGC~1068 \citep{kc00a,kc00b}. 
The fact that the soft X-ray emission is roughly coincident with the optical
emission-line gas in Markarian~3 \citep{sak00} is further evidence for
a heterogeneous NLR.

As noted in \S\ref{modelinput}, the central source in Markarian~3 is quite
luminous, which presents a problem for a simple, bi-conical model of
the emission-line gas. For example, [O~II] $\lambda$3727 is predicted
to be the strongest optical forbidden line for photoionized gas
characterized by $U \lesssim 10^{-3}$ \citep{fer83}.  
For that limit in $U$, our estimate for
$Q$ requires a density $n_{H}~>~10^{5.7}$~cm$^{-3}$ for a cloud at a
distance of 100 pc from the central source.  However, the critical
density for the $^{2}D_{3/2}$ level of O$^{+}$, the upper level of the
3726~\AA~ line, is 1.6 $\times$ 10$^{4}$ cm$^{-3}$ \citep{ost89}.
Therefore, lines from low ionization species, originating from levels
with low critical densities, will be collisionally suppressed since gas
directly exposed to the ionization radiation in the inner NLR of
Markarian~3 must have densities that significantly exceed the critical 
density.

The ionizing photon flux can be significantly reduced by an
optically-thick screen.  This effectively reduces the ionization state
of the component, but the emission line flux will be weak due to the
low emissivity of the gas.  The emission line flux can be increased by 
adding more emitting material to the new model component.  
However, while matching the observed emission lines, 
we found that the emitting area of 
the gas can exceed the geometrical limits imposed by the bi-cone 
geometry and the STIS slit. 
This constraint can be met if the new low ionization state component
lies further from the AGN than the other two model components.  This
gas would be outside the nominal bi-cone derived from the
$[$\ion{O}{3}$]$~$\lambda$5007 kinematic analysis \citep{rui01}.
Interestingly, \citet{kc00a} similarly found that the emission near
the NGC~1068 ``Hot Spot'' could be modeled by multiple components in
which the lowest ionization state component is not co-planar with the
others.

To test whether the lowest ionization state gas is more spatially
extended than the high and medium ionization state gas we compared the
$[$\ion{O}{3}$]$~$\lambda$4959 emission with that of
$[$\ion{O}{2}$]$~$\lambda$3727 in the STIS spectral images (since
these lines have similar fluxes). We extracted a
subimage centered on the $[$\ion{O}{3}$]$~$\lambda$4959 line from the
G430L spectral image.  The subimage size corresponds to a dispersion
of 4000~km~s$^{-1}$ at 4959\AA~ and 3$\farcs$0 in the cross-dispersion
direction.  We derived an average continuum value from the line-fits
described in Paper I and subtracted it from the subimage.  We
extracted a subimage centered on the $[$\ion{O}{2}$]$~$\lambda$3727
line with a size corresponding to the same velocity dispersion as the
$[$\ion{O}{3}$]$~$\lambda$4959 line subimage.  We subtracted the local
continuum from this subimage.  For direct comparison we used a linear
interpolation algorithm to transform the
$[$\ion{O}{2}$]$~$\lambda$3727 line subimage so that each pixel would
have the same velocity width as the $[$\ion{O}{3}$]$~$\lambda$4959
line subimage.

We show the $[$\ion{O}{2}$]$~$\lambda$3727 contours with
$[$\ion{O}{3}$]$~$\lambda$4959 contours overlaid in Figure~1.  Note
first the greater spatial extent of the $[$\ion{O}{2}$]$~$\lambda$3727
line ($\sim$2~$\arcsec$) compared with that of the
$[$\ion{O}{3}$]$~$\lambda$4959 emission ($\sim$1$\farcs$5).  The
$[$\ion{O}{2}$]$~$\lambda$3727 emission is more extended by five
spatial resolution elements in the West and 1 resolution element in
the East.  The $[$\ion{O}{2}$]$~$\lambda$3727 line also shows greater
extent in the velocity dispersion axis.  On the redshifted side this
line is more extended by at least a velocity resolution element, while
on the blueshifted side it is extended by up to one resolution
element.  However, we do not interpret the greater extent in the
dispersion direction as evidence that the lower ionization state gas
has significantly different kinematic behavior than the high and
medium state components comprising the nominal NLR bi-cone.  Instead,
this is most likely due to a difference in the spatial distribution of these
components within the slit in the dispersion direction.  For example,
the low ionization state gas may surround high and medium ionization
state gas as has recently been suggested for NGC~4151 \citep{kra08}.  
Therefore, the low ionization state gas may 
lie outside the nominal bi-cone configuration.

Another striking difference between the two spectral image
contour maps is in the 0$\farcs$0 measurement extraction bin.
The  $[$\ion{O}{3}$]$~$\lambda$4959 line shows a strong
blueshifted peak.  The $[$\ion{O}{2}$]$~$\lambda$3727 line
shows no corresponding feature at this position.  
The observed flux in the  $[$\ion{O}{3}$]$~$\lambda$4959 line is greater
than that in the $[$\ion{O}{2}$]$~$\lambda$3727 line 
in this measurement bin (cf. Table~2 in Paper I).
This contrast in line morphology may also be due to the more extended
structure of the low ionization state gas.

We compared the $[$\ion{O}{3}$]$~$\lambda$4959 spectral image
morphology to that of $[$\ion{Ne}{5}$]$~$\lambda$3426 to determine
whether there might be evidence for spatial differences between high
and medium ionization emission-line gas.  The contours for those lines
shown in Figure~2 show general correspondence for the major inner
contour concentrations, although the centroids of the concentrations
are slightly offset.  The outer contours of both lines on average span
the same range in the dispersion direction.  In the spatial direction
the 15\% contours for $[$\ion{Ne}{5}$]$~$\lambda$3426 do not extend
past 0$\farcs$4 east while those for $[$\ion{O}{3}$]$~$\lambda$4959
reach 0$\farcs$6 east.  This may indicate that there is less high
ionization state gas this far east along the STIS slit 
(see the discussion below in \S\ref{modelfits}). 
We find no evidence for spatial separation between the high
and medium ionization state gas from this data set.  
Hence, such components could be co-located within the bi-cone walls.

Although the $[$\ion{O}{2}$]$~$\lambda$3727 and
$[$\ion{O}{3}$]$~$\lambda$4959 lines show different spatial 
morphologies, we found the same number of 
kinematic components in each measurement bin
for the $[$\ion{O}{2}$]$~$\lambda$3727  profiles 
as we did for the $[$\ion{O}{3}$]$~$\lambda$5007
line.  The kinematic components in both lines show the same sign relative to 
the Markarian~3 systemic velocity, i.e., blueshifted components 
in $[$\ion{O}{3}$]$~$\lambda$5007 are blueshifted in 
$[$\ion{O}{2}$]$~$\lambda$3727.  Redshifted components in 
the two lines are likewise matched.  
The $[$\ion{O}{2}$]$~$\lambda$3727 emitting kinematic 
components may have slightly different velocities and
velocity dispersions than the $[$\ion{O}{3}$]$~$\lambda$5007 
emitting components.  However, these differences cannot be 
distinguished from this data set better than the velocity resolution 
of $\sim$316~km/sec at these wavelengths.  
Therefore, we maintain the use of the $[$\ion{O}{3}$]$~$\lambda$5007 
line kinematic parameters as templates for separating the 
blended kinematic components in other emission lines, including 
$[$\ion{O}{2}$]$~$\lambda$3727, as described in \S\ref{observations}.

\subsection{Model Components}\label{modelcomponents}
  
As we noted in the previous section the STIS spectrum of Markarian~3
NLR shows emission lines from ions spanning a wide range in ionization
potential and from energy levels spanning a wide range in 
critical density for collisional de-excitation.  
Furthermore we provided evidence that the 
$[$\ion{O}{2}$]$~$\lambda$3727 emission arises in a 
low ionization state component that 
is spatially distinct from the kinematically defined bi-cone.  
It is likely that the spectrum along any sight line to the NLR is
produced by a heterogeneous ensemble of gas components \citep{kra86} with 
different characteristic ionization 
parameters, hydrogen number densities and dust-to-gas ratios.  In order to
fit the observed line ratios and fluxes, we have included up to three
model components for each radial position along the slit.

A high ionization state component (labeled ``high'') produces
most of the line emission observed from high ionization potential
ionic species such as $[$\ion{Ne}{5}$]$~$\lambda$3426 and
$[$\ion{Fe}{7}$]$~$\lambda$6087.  The presence of [Fe VII] emission
is evidence that there is little depletion of iron onto dust grains
within the high ionization gas, hence we assumed that this component
is dust-free. We included a component with a lower ionization state
(labeled ``medium'') which would produce lines from lower ionization
potential species such as C~III] $\lambda$1909, [Ne~III]
$\lambda$3869 and $[$\ion{S}{3}$]$~$\lambda$9532. The
extinction-corrected Ly$\alpha$/H$\beta$ ratios (see Table 3 in Paper
I) are generally lower than predicted for Case B recombination
\citep{ost89}. This suggests that dust must be present in some
component of the NLR gas. Therefore, following \citet{kra86}, we
assumed that dust was mixed with the emission-line gas in the
``medium'' component. However, the presence of strong C~III]
$\lambda$1909 suggests that the dust/gas ratio is less than in the
ISM of the Galaxy (e.g. \citet{mat77}). Hence, we selected a 50\%
depletion factor onto grains for ``medium''.  The third component
(labeled ``low'') is situated outside the nominal bi-cone.  We did
not include dust in the ``low'' component, since that decreased the
emissivity of the gas.  This decreased emissivity  hampered our 
ability to reproduce the observed $[$\ion{O}{2}$]$~$\lambda$3727 flux. 
However, we cannot rule out the presence of some dust within this component, 
albeit at less than the ISM dust/gas ratio.

The model components may be either matter bounded 
or radiation bounded. \citet{bin96} described matter bounded clouds 
as fully ionized and ionization bounded (or in this paper ``radiation 
bounded'') clouds as partially ionized.   In Paper~I (\S5.2) we 
inferred the presence of a mixture of matter and radiation bounded 
clouds in the Markarian~3 NLR.   The component boundaries 
are model output parameters and are tabulated in \S4.2. 
 
\citet{ale99} argued that the NLR emission-line ratios in the Seyfert~1 
galaxy NGC~4151 indicated that the ionizing continuum was strongly
absorbed above the He~II Lyman limit. In a photoionization analysis of
{\it HST}/STIS spectra of NGC~4151, \citet{kra00} demonstrated that
the intervening absorbers resembled the intrinsic absorption detected
along the line-of-sight to the active nucleus. Since such an absorber
will reduce the He$^{+}$ ionizing photon flux at E$\gtrsim$ 54.4~eV,
the \ion{He}{2}~$\lambda$4686/H$\beta$ ratio will be lower than 
that predicted by photon-counting arguments.
Based on previous studies \citep[e.g.][]{kra00}, much of the NLR gas 
is matter bounded.  This has the effect of increasing 
\ion{He}{2}~$\lambda$4686/H$\beta$, since the H$^+$/H$^0$ transition 
zone may not be present.  Therefore, in order to reproduce the observed 
\ion{He}{2}~$\lambda$4686/H$\beta$,  
it is plausible 
that the intrinsic continuum is absorbed above the \ion{He}{2} Lyman limit.
Based on our
preliminary modeling, we found that an absorber with column density
10$^{20.4}$~cm$^{-2}$ and ionization parameter 10$^{-1.5}$ yielded the
best fits for these lines ratios (see Figure~3).  Since the absorber
is required for all of the observed components it must be well within
the size of a detector spatial resolution element (26~pc).  

As discussed above, the ionizing radiation to which the ``low''
component is exposed must be heavily filtered by gas close to the
central source. In order to determine the characteristics of the
screening material, we created a grid of screen models with a range in
ionization parameter and in column density that absorb more ionizing
photons than the screen for the ``high'' and ``medium'' ionization
state components.  The screens were sorted by transmitted photon flux.
We selected the screen that appropriately reduced the ionizing
continuum flux and yielded the best matches to the data for the lines
$[$\ion{O}{2}$]$~$\lambda$3727 and
$[$\ion{N}{2}$]$~$\lambda\lambda$6548,6583 relative to H$\beta$ from
the emitting cloud of interest.  We used three types of absorbers for
the low ionization state model components: (a) $U$~=~10$^{-2.5}$ and
$N_{c}$~=~10$^{20.7}$~cm$^{-2}$, (b) $U$~=~10$^{-1.5}$ and
$N_{c}$~=~10$^{21.6}$~cm$^{-2}$ and (c) $U$~=~10$^{-3.0}$ and
$N_{c}$~=~10$^{21.9}$~cm$^{-2}$.  The screened continua produced by
each of these absorbers are shown in Figure~3.  These
screens absorb nearly all of emitted continuum from 13.6~eV to 200~eV.
Most of the low ionization state clouds were screened by the
$N_{c}$~=~10$^{21.6}$~cm$^{-2}$ absorber.  However, the eastern-most
low ionization state cloud was screened by the
$N_{c}$~=~10$^{20.7}$~cm$^{-2}$ absorber while both the blueshifted
and redshifted low ionization state clouds in the adjacent 0$\farcs$3
measurement bin were screened by the $N_{c}$~=~10$^{21.9}$~cm$^{-2}$
absorber.

Interestingly, these absorbers have physical parameters similar to the
intrinsic absorbers detected in some Seyfert 1 galaxies.  The low
column density absorber that screens the nominal bi-cone gas is
similar to the UV absorbers detected in NGC~5548 \citep{cre03}. The
screens for the low ionization state gas have column densities
intermediate between those of the UV and X-ray absorbers in NGC~4151
\citep{kra01}.

\section{Model Results}\label{modelresults}

\subsection{Revised View of the NLR Structure and Final Model Input Parameters}\label{nlrstructure2}

Based on our preliminary analysis, incorporating 1) the \citet{rui01}
kinematics study and 2) the evidence for morphological differences in
emission line spectral images from low and high ionization potential
ionic species, we developed a revised model for 
the NLR structure, shown in Figure~4.  
Without higher spatial and spectral resolution observations of the 
$[$\ion{O}{2}$]$~$\lambda$3727 line emission it is difficult 
to constrain the location of the low ionization components. 
As a guide for the models we located each low
ionization component on the observer's sight line midway
between the outer edge of the NLR bi-cone and a line that sweeps out
two solid angles (East and West) of $\pi$~steradians each when rotated
around the bi-cone axis.  This line is at a 38.7$\degr$ angle with
respect to the bi-cone axis.  
The ionizing continuum for the low
ionization region (lighter grey shaded area in the figure) is more
heavily absorbed than that illuminating the high and medium ionization
regions.  As noted, the screening is non-uniform for the low
ionization state gas.

Although we have relaxed our constraint that all the emitting 
gas lie within the walls of the kinematically derived hollow bi-cone, 
we maintained the geometrical constraints for the high and low 
ionization state components.  Those components produce 
nearly all of the $[$\ion{O}{3}$]$~$\lambda$5007 emission 
on which the kinematically derived NLR structure is based. 
We loosely applied  similar geometrical constraints on the 
low ionization state components using the inner and outer 
limits described above. 

We list the final CLOUDY input parameters for all
modeled components corresponding to our measured kinematic
components in  Table~1 of Appendix~\ref{tables}.
We did not include a dusty medium ionization
state component for the central blue-shifted
kinematic component, bin (0$\farcs$0E~(b), 0$\farcs$3), since 
a two-component (high+low) model adequately reproduced
the observed emission spectrum for this position in the NLR.
In Figures~5 through 7 
show the hydrogen number density 
($n_{H}$) as a function of radial
distance from the AGN for the high, medium and low ionization
state components, respectively.   Recall that we
constrained the NLR density such that clouds further
from the AGN have lower density that those closer to the AGN 
(see details in Paper~I). 
However, we assumed no specific
functional form for the cloud volume number density vs. position.
For the high ionization parameter model components, the
$n_{H}$ decreases faster than $r^{-2}$
in both the East and West along the STIS slit.  The $n_{H}$
within the dusty medium ionization state component
falls off as approximately $r^{-2}$  in the West.
It is difficult to draw any conclusion about the
radial dependence of the cloud hydrogen density
in the east since there are only three data points for this component.
The low ionization state component shows a density decrease
in the West consistent with $r^{-2}$.  In the East,
the data points follow the same functional form as those
in the West out to $r~\sim~$150~pc.
For comparison, \citet{kra00} found the hydrogen density proportional to
$\sim~r^{-1.6}$ in the southwest along the STIS slit
and $\sim~r^{-1.7}$ in northeast
for all  model cloud components in
the NGC~4151 NLR.

\subsection{Fit to the Observations}\label{modelfits}

In  Tables~2 through 12 of Appendix~\ref{tables}
we list the line fluxes relative to H$\beta$
for each model component, the composite model, and for the observed
data.  The good agreement between the model emission line ratios and
the observed emission line ratios indicates that our main assumptions
about the NLR structure and intrinsic ionizing continuum were
reasonable.  Furthermore, the ionizing radiation emitted by the
central source is sufficient to power the NLR, without additional
ionization mechanisms such as shock-heating or starbursts.  As
suggested for other Seyfert galaxies
\citep[e.g.,][]{kra00,kc00a,kc00b},
multiple components with a range in ionization states comprise the NLR
gas.  Although the model predictions are consistent with at least some
dust mixed in with the emission-line gas, the dust/gas ratio appears
to be substantially less than that in the Galactic ISM.  However,
perhaps the most striking prediction of these models is that there is
a strong contribution from a low ionization gas component which lies
outside the nominal bi-cone structure. This provides new insight into
the nature of the circumnuclear gas close to the AGN, which we will
discuss in \S\ref{collimation}.

In Paper I we suggested that the continuum might be more heavily
absorbed in the East than in the West based on the relative weakness
of the high ionization potential emission lines
$[$\ion{Ne}{5}$]$~$\lambda$3426 and
$[$\ion{Fe}{7}$]$~$\lambda\lambda$5722,6087 on the eastern side of the
NLR.  However, we were able to fit these lines using a uniform screen
in both directions for the high ionization state model component from
which these observed lines would originate.  Indeed the absorbing
screen diagnostic ratio \ion{He}{2}~$\lambda$4686/H$\beta$ shows no
strong East/West asymmetry.  In the models the effective ionization
parameter and number density of the high ionization state component
show no obvious asymmetry or trends with direction either.  However, 
the high ionization state model components' emitting areas and masses are 
greater in the west than in the east.  Perhaps the distribution 
of the high ionization state gas is asymmetric.  

We list additional CLOUDY model output values, derived parameters and
geometrical constraints in Table~13 of Appendix~\ref{tables}.  
The emitting area, component
depth, bin depth, component height and bin height were defined in
\S\ref{nlrstructure2}.  The ratios of component to bin depths for 
the high and medium
ionization state model components meet the geometrical constraint
(upper limit of 1.5).  The component heights for these components are
much less than their corresponding measurement bin heights.  Within
each measurement bin the sum of the volume filling factors for the
high and medium ionization components is less than unity.  Therefore,
it is unlikely that clouds at low radial distance from the AGN within
the kinematically defined bi-cone shield clouds at higher radial
distances.   

Three of the low ionization components have 
component depth to bin depth ratios that exceed 
the 1.5 upper limit:  (0$\farcs$3E, r, 0$\farcs$3), 
(-0$\farcs$7W, r, 0$\farcs$3) and (-1$\farcs$0W, r, 0$\farcs$3).
The second of these three has a filling factor near unity.  
It is possible this component may shield  NLR gas 
located further from the AGN (measurement bin -1$\farcs$0W, r, 0$\farcs$3), 
although we did not explore this possibility.

We created alternate models for the (-0$\farcs$7W, r, 0$\farcs$3) 
component.  In these models the  number density 
ranged from 10$^{2.0}$~cm$^{-3}$ to 10$^{2.5}$~cm$^{-3}$ and 
the column density ranged from 10$^{19}$~cm$^{-3}$ to 10$^{23}$~cm$^{-3}$. 
All models with good line fits violated one or another geometrical 
constraint.  
However, the geometry of the low ionization state gas outside the 
kinematically defined bi-cone is not well constrained. 
We conclude that more low-ionization gas is required 
to match the observed line ratios than would fit within the 
geometrical guidelines selected for the gas outside the nominal bi-cone. 
The extent of the low ionization gas along our lines of sight 
in the (0$\farcs$3E, r, 0$\farcs$3), 
(-0$\farcs$7W, r, 0$\farcs$3) and (-1$\farcs$0W, r, 0$\farcs$3)
measurement bins likely exceeds our geometrical guidelines for 
these components. 
It is possible that the lines of sight in the measurement bins 
(0$\farcs$3E, r, 0$\farcs$3), 
(-0$\farcs$7W, r, 0$\farcs$3) and (-1$\farcs$0W, r, 0$\farcs$3)
intersect foreground or background extended narrow-line region (ENLR) gas. 

Based on the predicted electron temperatures (Table~13,
Appendix~\ref{tables}) and densities (Tables 2 through 12,
Appendix~\ref{tables}), it is apparent that the co-located ``high''
and ``medium'' components are not generally in pressure equilibrium. A
more highly ionized medium, with a large volume filling factor (see
\S\ref{model2xray}), could confine one or both of these components.
The apparent drop in density as a function of radial distance suggests
that the UV/optical emission-line gas is not fully confined.  The
clouds may be part of an outflow originating close to the AGN and they
may expand as they traverse the NLR \citep[see][]{rui01}.  
However, it is likely that the clouds are at least partially confined, otherwise 
the density would decrease more rapidly with distance than our models predict.

Although the model predictions provide a good fit to the data,
some discrepancies are evident.
In several positions, notably 0$\farcs$3E~(r), 
0$\farcs$0W~(r) and -1$\farcs$0W~(r), some combination 
of one, two or all three of the set of lines 
\ion{C}{3}$]$~$\lambda$1909, \ion{C}{2}$]$~$\lambda$2326 
and  $[$\ion{Ne}{4}$]$~$\lambda$2424 are underestimated. 
Recall that we used the LMC extinction curve 
of \citet{koo81} to correct our observations for 
intrinsic reddening within Markarian~3.  This curve has a 
2200\AA~ bump similar to that in the Galactic 
curve of \citet{sav79} on the long wavelength 
side of the feature.  Toward shorter wavelengths 
the \citet{koo81} extinction curve rises faster than 
the \citet{sav79} curve.  This may be too steep 
for Markarian~3.  If this is so, then we 
have over-corrected the line fluxes 
for extinction at these wavelengths, although   
the models do not consistently underestimate 
all three lines.   Note that these three kinematic components 
are redshifted and therefore lie on the far side of the galaxy 
plane (at positions 0$\farcs$3E~(r) and 0$\farcs$0W~(r)) 
or in the galaxy plane (position -1$\farcs$0W~(r)) (see Figure~4) 
The corrected line fluxes at these positions may be more sensitive to 
uncertainties in the extinction correction than those at other positions.  
In particular, the positions 0$\farcs$3E~(r) and 0$\farcs$0W~(r) 
are in the region along the STIS slit that shows the greatest  
reddening (see Figure~8 in Paper~I).  

The model line fits for the position 0$\farcs$5E~(r) did not 
meet the criterion that half the model lines match the data 
within $\pm$30\%.  Generally the line fluxes in this 
component are underestimated by the model.   If the 
column density of the ``medium'' ionization state model 
component is increased, improved fits to the data can be 
achieved;  however, the emitting area will then violate 
our geometrical constraint for this measurement bin.  
This position is the most heavily reddened along the STIS slit and 
the UV lines shortward of 2400\AA~ are poorly fit.  
It is, thus, possible that the poor fits are due to over-correcting the 
measured line fluxes for extinction as discussed above. 
The underestimated \ion{He}{2}~$\lambda$4686 model line flux 
may be an indication that the intrinsic absorption screen 
is too thick for this measurement position.  

$[$\ion{S}{3}$]$~$\lambda$9532 is under predicted in the bins
0$\farcs$3E~(r), 0$\farcs$0E~(b), 0$\farcs$0W~(r), -0$\farcs$3W~(b)
and -0$\farcs$W~(r).  Again this may be due to uncertainty in the
shape of the extinction curve, and in this case the long base-line
between the emitted line wavelength and the \ion{He}{2}~$\lambda$4686
extinction reference line wavelength.

The predicted Ly$\alpha$~$\lambda$1216/
H$\beta$~$\lambda$4861 ratio is low at positions 
0$\farcs$5E~(r), 0$\farcs$3E~(r), 0$\farcs$0W~(r), 
-0$\farcs$7W~(r), -1$\farcs$0W~(r).  As described above 
the extinction curve used to correct these observations 
\citep{koo81} is steeper than the Galactic curve 
\citep{sav79} at wavelengths shorter than 2200\AA. 
The actual extinction curve required for Markarian~3 
may lie somewhere between these two. 

Two positions, 0$\farcs$3E~(b) and 0$\farcs$0E~(b) 
show overestimated \ion{C}{4}~$\lambda\lambda$1548,1551 flux. 
It is likely that 
the measurements for these lines at this position 
are poor, although this is not reflected in the 
$\pm$1$\sigma$ error bars.  Even though the  flux ratio 
of the doublet is constrained (2:1), it is difficult to separate 
the four (2 doublet $\times$ 2 kinematic component) lines. 
The \ion{Mg}{2} flux is overestimated at position 
0$\farcs$0E~(b).  This is also likely due to poor separation 
of the blended doublet/kinematic component lines.

\subsection{Model Comparison with X-Ray Data}\label{model2xray}

We compared our model estimated X-ray lines with those observed by
\citet{sak00} with $Chandra$/HETGS.  They obtained a spectrum from
0.5~$<~E~<$~10~keV (or 1~$<$~$\lambda$~$<$~24\AA) through the
11$\arcsec$~$\times$~19$\arcsec$ aperture.  The aperture was oriented
with the the cross-dispersion (longer) axis at position angle
90$\degr$ to measure the flux from the entire NLR~+~ENLR.  They
detected eighteen resonance lines from the highly ionized H-like and
He-like ionic species of O, Ne, Mg, Si and S.  They also detected ten
lines from Fe$^{+20}$ through Fe$^{+25}$. They suggested that the
X-ray emission-line gas was photoionized by the central source, thus
it is useful to determine what, if any, contribution the UV/optical
emission-line gas may have to the X-ray spectrum.

For each X-ray line we summed the fluxes computed by CLOUDY for all of
our model components.  To compare our predicted lines with the
\citet{sak00} observations we must normalize the predictions using one
of the lines in common.  We assumed that the O$^{+6}$ gas has the same
distribution as the H$^{0}$ gas and used the \ion{O}{7}~$\lambda$22.10
line for normalization.  We found a predicted flux of
$\sim$10$^{-14}$~ergs~s$^{-1}$~cm$^{-2}$.  This is 30\% of the
observed flux for this line.  The estimate is reasonable considering
the difference in aperture sizes between STIS and HETGS.  Our models
predict significant emission only from the other oxygen lines
observed: the predicted
\ion{O}{7}~$\lambda$21.60/\ion{O}{7}~$\lambda$22.10 ratio is 0.5 and
the \ion{O}{7}~$\lambda$21.81/\ion{O}{7}~$\lambda$22.10 ratio is 0.7.
These ratios are somewhat higher than those reported in \citet{sak00}.
The physical conditions (e.g., optical depth and/or microturbulence)
may be different in the the O~VII emitting gas which lacks a strong
UV/optical footprint.
Also, since the models underestimate the observed flux for lines from
the higher ionization potential ionic species, it is likely that the
Markarian~3 NLR contains a yet higher ionization component, which
produces little or no UV/optical emission-line flux. Given the low
volume filling factors predicted for the UV/optical emission-line
clouds (see Table~13 in Appendix~\ref{tables}), 
and the weak attenuation of the ionizing
continuum expected from the highly ionized gas, the X-ray emitters
could be co-located with the ``high'' and ``medium'' components
without any obvious affect. In this case, the volume filling factors
for the X-ray emitters would far exceed those of the UV/optical knots,
perhaps filling the bi-cone envelope.

\section{Collimation of the Ionizing Radiation}\label{collimation}

Based on our models, the ionizing photon flux (and hence the
gas ionization state) in the NLR decreases
with increasing polar angle from the symmetry axis
(nearly coincident with the sky plane).
We suggest that the collimation of the ionizing radiation
is not sharply defined by an opaque molecular torus
as described in most unified model scenarios.
Some alternative mechanisms for collimation are
a ``torus atmosphere'' or an accretion disk wind.

\citet{eva93} proposed a torus atmosphere model for NGC~4151
with a column density of order 10$^{20}$--10$^{21}$~cm$^{-2}$
that absorbed X-rays but transmitted non-ionizing UV/optical
radiation. Another model suggested by \citet{eva93}
was a clumpy torus composed of clouds with a range of
column densities and spatial distributions fortuitously
arranged to allow transmission of the optical BLR and continuum emission
while absorbing X-ray emission.
\citet{fel99} suggested that the intrinsic X-ray absorption
detected in the Seyfert 1 galaxy Marakarian~6
arose in a similar torus ``atmosphere''
of larger column density (10$^{22}$~cm$^{-2}$).
% Interestingly, both of these galaxies possess a linearly extended
% NLR \citep{sch96,sch03}, similar to that 
% observed in Markarian~3, which suggests a similar
% orientation. 
Interestingly, the narrow line region in each of these galaxies 
appears as a linear (as opposed to circular) feature in 
$[$\ion{O}{3}$]$ images \citep{sch96,sch03}.  This morphological 
similarity with Markarian~3 suggests that these three 
NLRs share a common orientation. 

In the hydromagnetic wind model of \citet{kon94}
an outflow is driven by the angular momentum lost
by matter in the an accretion disk.   The outflow is
stratified such that  density increases with distance from
the nucleus.
Electrons in the wind collimate the ionizing radiation
which decreases with increasing polar angle.
The gas close to the symmetry axis is highly ionized and 
the ionization state decreases with radius and polar angle.

The gas ionization state and density gradients are consistent with our
picture of the circumnuclear gas in Markarian~3 developed from the
analysis of the NLR emission.  However, the ionizing radiation in our
models is collimated through absorption by the circumnuclear gas
instead of electron scattering.  In the \citet{kon94} model, the outer
regions of the wind at high polar angle may contain dust.  This dust
may obscure the AGN as required by the unified model, but it is not
considered a physically distinct opaque component such as the putative
torus.  We did not include dust in our model screens.  The screens may
lie within the dust sublimation radius \citep[see][]{bar87} or they may
be part of a wind that originates in a dust-free region.  However, the
physical conditions in the screening gas are not well constrained, so
we cannot rule out the possibility that they contain dust. The main
point is that the characteristics of the NLR are consistent with
collimation of the ionizing radiation by ionized absorbers.
Moreover, Markarin~3 may not be a unique case. \citet{kra08}
used $[$\ion{O}{3}$]$ and $[$\ion{O}{2}$]$ images obtained with 
{\it HST}/WFPC2 to map the ionization
structure of the NLR in NGC~4151. They found that the ionization state of the
of NLR, determined via the $[$\ion{O}{3}$]$/$[$\ion{O}{2}$]$  ratio, 
dropped with increasing
distance from the bicone axis, which is also consistent by a collimation
of the ionization radiation by gas near the AGN.

\section{Black Hole Mass, Accretion Rate and Outflow}\label{bhmass}

\citet{woo02} derived a black hole mass of 
$\approx4.5~\times~10^{8}~M_\odot$ 
for Markarian~3 based on the $M_{BH}$ - stellar velocity
dispersion relationship.  Using the bolometric luminosity of the model
input continuum, 2~$\times$~10$^{45}$~ergs~s$^{-1}$, we estimate that
the AGN is radiating at $\sim$3.5\% of its Eddington limit.
\citet{kra04} and \citet{pet04} show evidence that many AGN radiate at
$\sim$10\% of their Eddington limits.

We estimated the mass of the NLR gas using the volume and number
density values for the clouds (see Table~13 in Appendix~\ref{tables}), 
obtaining a value of 
$2~\times~10^{6}~M_{\odot}$ for the small portion of the NLR gas modeled
in this study. By comparison, the NLR mass estimated for NGC~4151 from
the photoionization models by \citet{kra00} is $\sim 4 \times 10^{3}
M_\odot$.  This would correspond to $\sim 2 \times 10^{4} M_\odot$ for
the larger region sampled in Markarian~3.  We estimate that we have
sampled and modeled one-sixth of the hollow bi-cone volume with a
400~pc extent along the bi-cone axis.  Assuming that the NLR mass
distribution is roughly azimuthally symmetric, the total NLR mass
along the 400~pc extent is $1.2 \times 10^{7} M_\odot$.  This is on
the order of the mass of a late-type dwarf galaxy.  For comparison, a
typical late-type dwarf galaxy, such as the LMC, has an atomic
hydrogen mass on the order of 10$^{8}$ $M_\odot$ \citep{swa02,sta03},
while the dwarf elliptical galaxy NGC~205 has an \ion{H}{1} mass of
approximately 10$^{7}$ $M_\odot$ \citep{wel98}.  \citet{noo05} found
an \ion{H}{1} bridge between Markarian~3 and UGC~3422, a type SAB(rs)b
galaxy $\sim$~100~kpc to the north-west, in the Westerbork \ion{H}{1}
Spiral and Irregular Galaxy Survey (WHISP).  This interaction is
likely responsible for the large amount of emission-line gas inferred 
by the models and for triggering and sustaining the level of 
activity observed in Markarian~3.

The mass accretion rate, $\dot M$, is determined from the time
derivative of Einstein's mass-energy equivalence relation and by
assuming an efficiency, $\eta$, for the conversion of matter to light
\begin{equation}
\dot M = \frac{L}{\eta c^{2} }
\end{equation}
Based on our estimated bolometric luminosity, and assuming
$\eta~=~0.1$, the accretion rate is 0.35~$M_\odot$ yr$^{-1}$.  This
can be compared to the mass outflow rate that we derived using the
cloud masses from the models, their distances from the AGN and their
radial velocities relative to the AGN.  We converted the line of sight
cloud velocities to outflow velocities relative to the central source
using the bi-cone model geometry. We obtain a mass outflow rate of
$\sim$15~$M_\odot$~yr$^{-1}$, or 42 times the accretion rate.  
This suggests that most of the infalling
material is blown out before it can be accreted by the central source 
at the epoch of these observations.
We find that the observed NLR outflow kinetic energy is
2~$\times$~10$^{55}$~ergs.  The kinetic energy luminosity,
3~$\times$~10$^{42}$~ergs~s$^{-1}$, is a small fraction of the
bolometric luminosity. Note that these estimates are lower limits for
Markarian~3 since we modeled only a portion of the entire NLR.

\section{Summary}\label{summary}

We have examined the physical conditions in the NLR of the 
Seyfert 2 galaxy Markarian~3, using low-resolution, long-slit
data obtained with {\it HST}/STIS (see Paper I)
and photoionization models. The main results of our 
photoionization modeling analysis are as follows:

1. We have shown that the Markarian~3 NLR UV/optical
emission spectrum can be modeled using photoionization
as the sole excitation mechanism. The bulk of the emission-line
gas lies within the envelope of the bi-conical region described
in the kinematic model of \citet{rui01}. We modeled the emission
from this region using two components, labeled ``high'' and ``medium''
to describe their relative ionization state. We determined
that the ionizing continuum incident upon this gas is best
modeled as a broken power-law (see Paper I), filtered through
a layer of intervening gas, closer to the AGN, that causes some
attenuation above the He~II Lyman limit. 
 
2. There is a third component of emission (``low'') in which lines
from levels with low critical densities from low ionization species,
e.g. [O~II] $\lambda$3727, arise. In order for such a component to
exist in the inner NLR of Markarian~3, it must be heavily shielded
from the ionizing source, presumably by an optically thick intervening
absorber.  Given the nature of the shielding and the low emissivity of
low density gas irradiating in such a manner, the ``low'' component
must lie outside the nominal emission bi-cone. The different
morphologies of the [O~II] $\lambda$3727 and [O~III] $\lambda$4959
lines are strong evidence for such a scenario.

3. The model parameters used to describe the intervening absorbers are
similar to those derived from photoionization studies of intrinsic UV
and X-ray absorbers in Seyfert 1 galaxies \citep{kra01,cre03}. This
suggests that intrinsic absorbers are responsible for the collimation
of the ionizing continuum, rather than the thick molecular torus
described by unified models \citep{ant93}.

4. The volume filling of the emission-line components are generally
small ($<$0.01), and the clouds of different ionization states are not
in pressure equilibrium.  The cloud densities decrease with radial
distance, which suggests that they are not fully
confined. Interestingly, our models predicted only 30\% of the
\ion{O}{7}~$\lambda$22.10 flux reported by \citet{sak00}\ and
negligible fractions of the observed flux for higher ionization
states. The additional soft X-ray emission-line gas could lie between
the bi-cone walls, and may partially confine the UV/optical clouds.

5. We found that the ionizing radiation could be produced by accretion
onto a black hole of $M_{BH} > 10^{8} M_\odot$, with the system
radiating at $<$ 10\% of its Eddington luminosity. The mass outflow
rate exceeds the inferred accretion rate by a factor of $\sim$ 40.

6. The large amount of NLR gas is consistent with that in a dwarf
elliptical galaxy.  This mass, the high luminosity of the ionizing
continuum and the dust structure (described in Paper I) may be
indications of a recent merger or fueling event.  Indeed, the \ion{H}{1}
map of \citet{noo05} shows evidence of such an interaction with the
neighboring spiral galaxy UGC~3422.

\acknowledgments

We thank Jane Turner for useful comments.
We thank Gary Ferland for the use of CLOUDY and for 
helpful advice on running the models. 
We thank Henrique Schmitt for calling our attention 
to the \ion{H}{1} map of \citet{noo05}.  
We thank the referee for constructive comments. 
Some of the data presented in this paper were obtained from the
Multimission Archive at the Space Telescope Science Institute (MAST).
STScI is operated by the Association of Universities for
Research in Astronomy, Inc., under NASA contract NAS5-26555. Support
for MAST for non-HST data is provided by the NASA Office of Space
Science via grant NAG5-7584 and by other grants and contracts. This
research has made use of NASA's Astrophysics Data System. 
This research has made use of the
NASA/IPAC Extra-galactic Database operated by the Jet Propulsion
Laboratory, and NASA's Astrophysics Data System Bibliographic Services.

We acknowledge the financial support of NAG5-4103 and 
NAG5-13109. 

Facilities: \facility{HST(STIS)}

\clearpage

\appendix

\section{Tables}\label{tables}

All tables should appear in this appendix. 

\clearpage

\placetable{1}
\placetable{2}
\placetable{3}
\placetable{4}
\placetable{5}
\placetable{6}
\placetable{7}
\placetable{8}
\placetable{9}
\placetable{10}
\placetable{11}
\placetable{12}
\placetable{13}

\clearpage

\figcaption[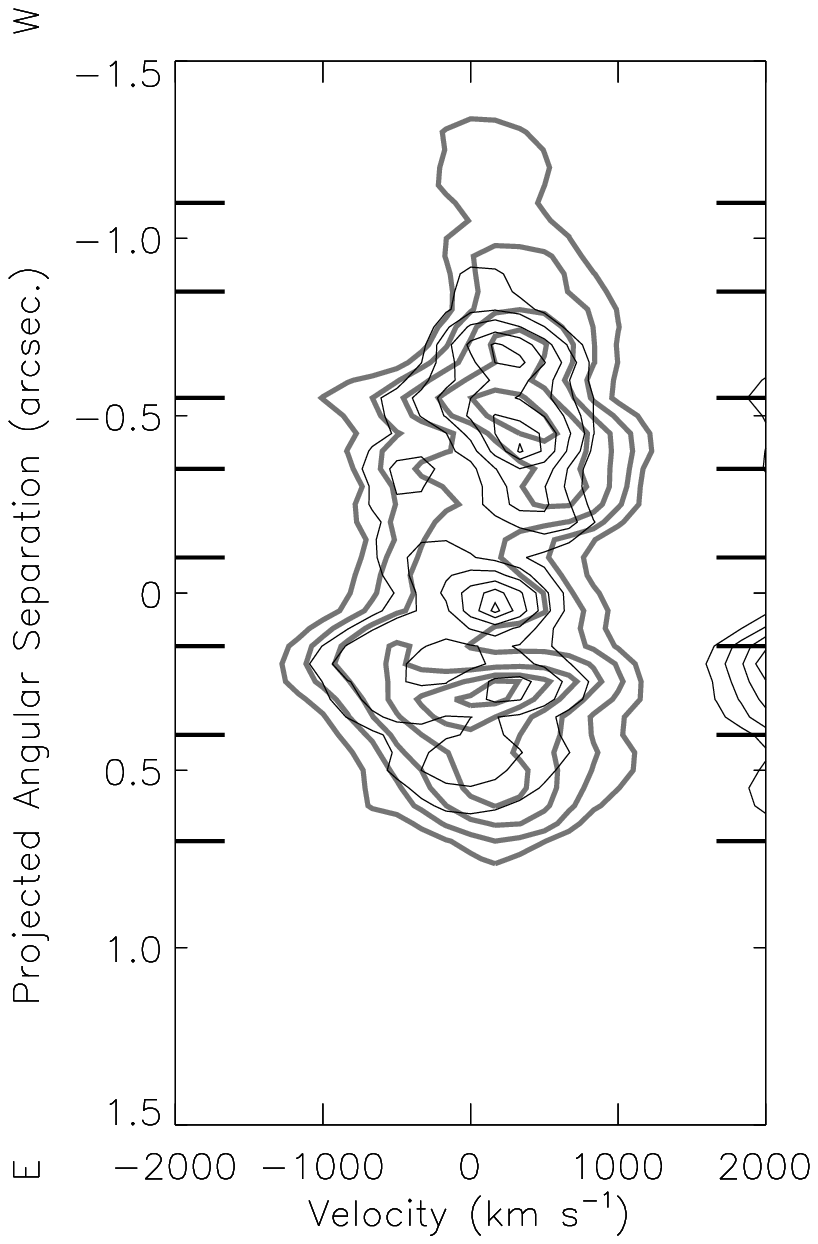]{The structure of [O~II] $\lambda$3727 (thick grey lines) emission compared
to [O~III] $\lambda$ 4959 (thin black lines). The thin black lines
near velocity +2000~km~s$^{-1}$ are contours from the neighboring
$[$\ion{O}{3}$]$~$\lambda$5007 line. The contours are the
15\%, 25\%, 40\%, 60\%, 80\%, and 95\% flux levels relative to the
peak value for the line of interest.
The velocity resolution is two detector elements or 316~km~s$^{-1}$.
The spatial resolution is 0$\farcs$1.
The heavy black tick marks
on the y-axis indicate the limits of the measurement bins.}

\figcaption[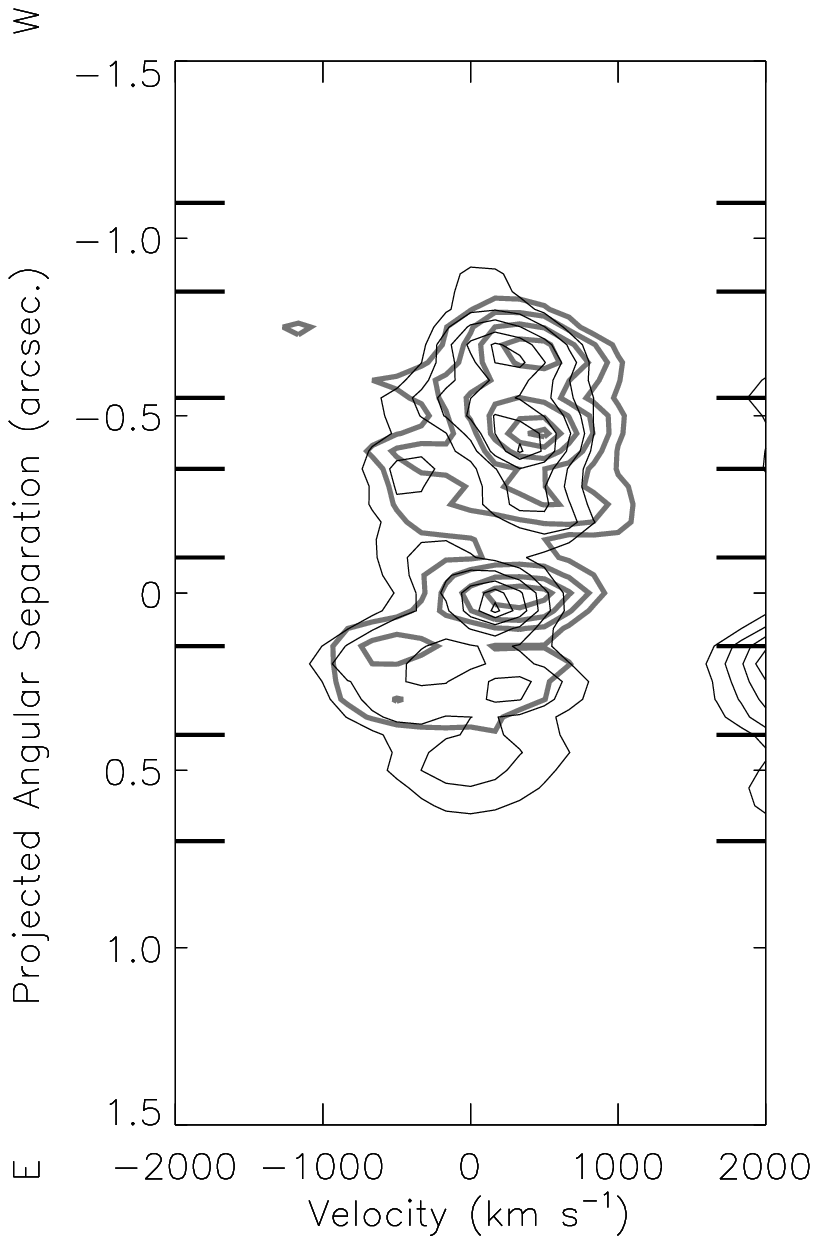]{The structure of [Ne~V] $\lambda$3426 (thick grey lines) 
emission compared to [O~III] $\lambda$ 4959 (thin black lines). 
The details of the plot are as described in the caption for Figure 1.}

\figcaption[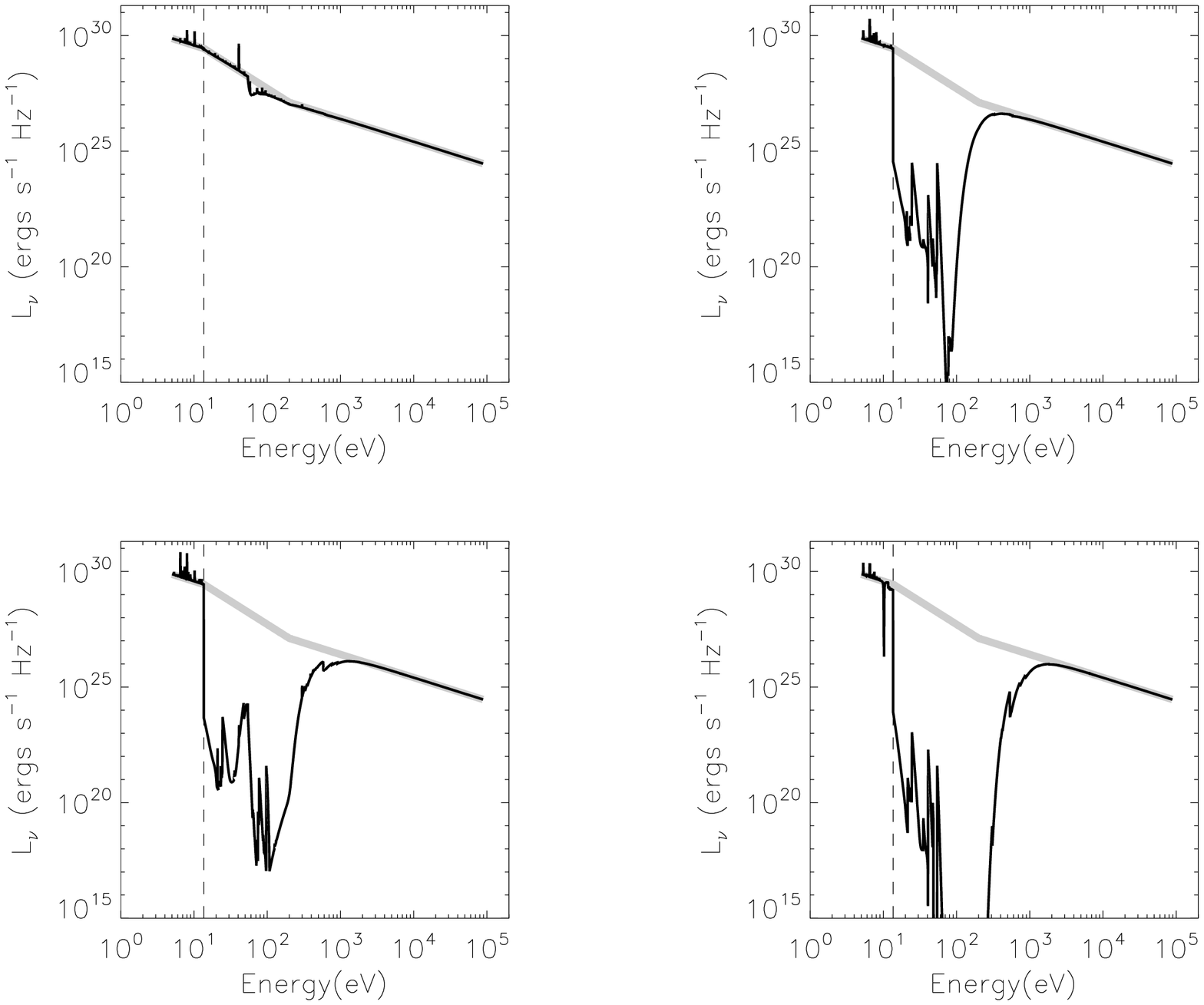]{The effects of various absorbers 
on the model intrinsic continuum (grey line in all panels) are shown.  
Top left:  filtered continuum (black line) illuminating gas within nominal 
bi-cone; absorber properties are $N_{H}$~=~10$^{20.4}$~cm$^{-2}$ and 
log$_{10}U$~=~-1.5.  
Top right: filtered continuum (absorber properties:   
$N_{H}$~=~10$^{20.7}$~cm$^{-2}$ and log$_{10}U$~=~-2.5)
illuminating gas outside the nominal bi-cone envelope. 
Bottom left: filtered continuum (absorber properties:   
$N_{H}$~=~10$^{21.6}$~cm$^{-2}$ and log$_{10}U$~=~-1.5)
illuminating gas outside the nominal bi-cone envelope. 
Bottom right: filtered continuum (absorber properties:  
$N_{H}$~=~10$^{21.9}$~cm$^{-2}$ and log$_{10}U$~=~-3.0)
illuminating gas outside the nominal bi-cone envelope.}

\figcaption[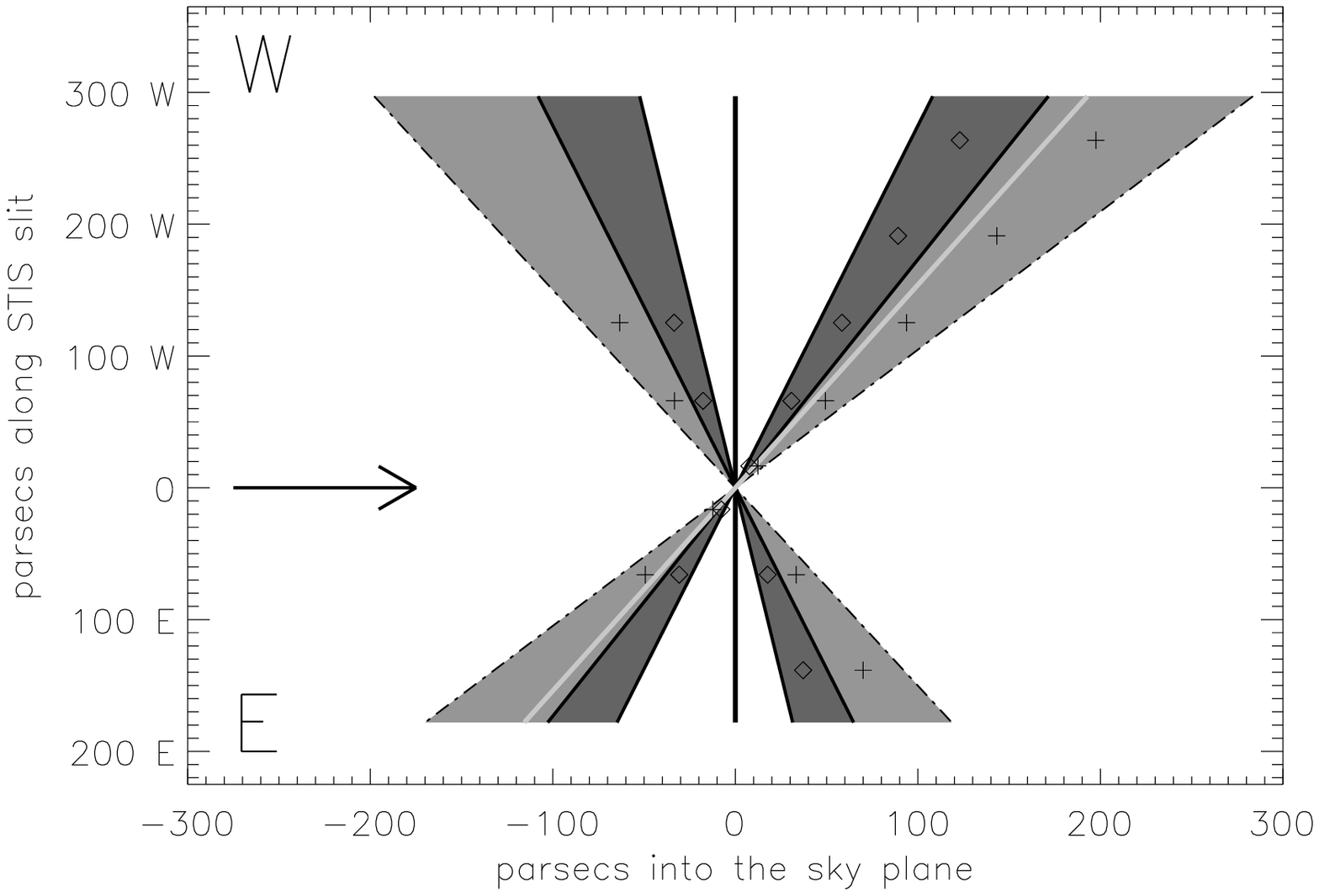]{In this schematic diagram the arrow indicates 
the observer's line of sight.  The vertical black line is the sky plane.
The darker grey bi-cone cross section indicates the nominal hollow
bi-cone of \citet{rui01}.  The diamonds in this region indicate the
distances from the AGN used in the CLOUDY models for the co-located
high and medium state ionization components that correspond to the
measured kinematic components.  The lighter grey region indicates the
location of the low ionization state gas outside the bi-cone.  The
crosses indicate the distances from the AGN used in the CLOUDY models
for the low ionization state components.  Redshifted kinematic components are 
to the right of the sky plane, and blueshifted components are to the left. 
The thick grey line shows the orientation of the host galaxy plane 
(see Paper I and references therein).}

% \figcaption[f5.eps]{Logarithm of the hydrogen volume number density	 
% for model clouds as a function of radial distance from the AGN.	 
% Density values for high, medium, and low ionization state components	 
% are shown in the left, center, and right panels, respectively.	 
% Redshifted components are shown in as open squares while blueshifted	 
% components are represented by solid squares.  The horizontal bars	 
% represent the deprojected widths of the measurement bins.  The solid	 
% curves represent functional forms for the hydrogen density vs. radial	 
% distance from the AGN.  The curve labels indicate the power of (1/$r$) 
% for the density dependence.}                                           

\figcaption[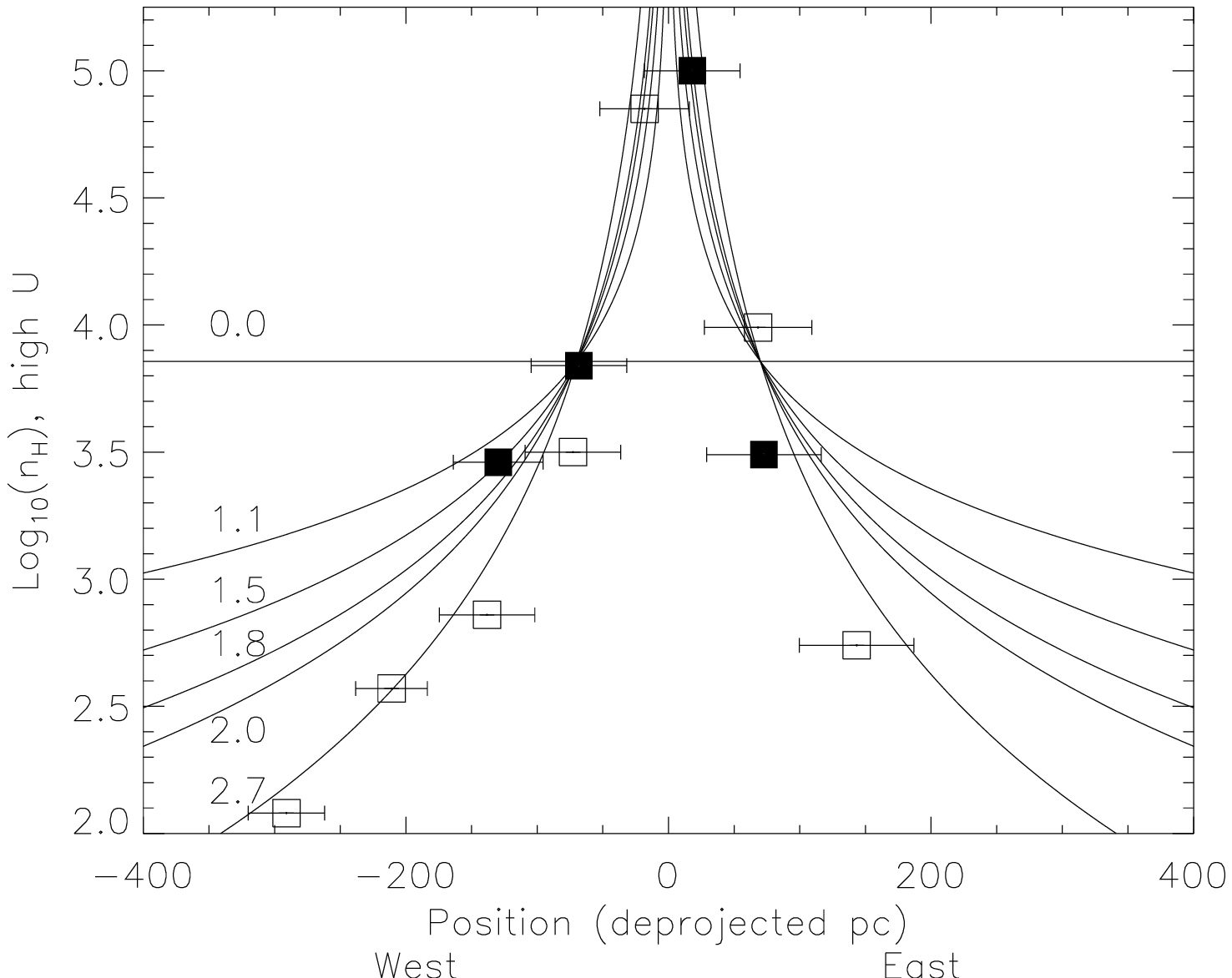]{Logarithm of the hydrogen volume number density	 
for high ionization state model clouds as a function of radial distance from the AGN.	 
Redshifted components are shown as open squares while blueshifted	 
components are represented by solid squares.  The horizontal bars	 
%represent the deprojected widths of the measurement bins.  The solid	 
represent the deprojected heights of the measurement bins.  The solid	 
curves represent functional forms for the hydrogen density vs. radial	 
distance from the AGN.  The curve labels indicate the power of (1/$r$) 
for the density dependence.}                                           

\figcaption[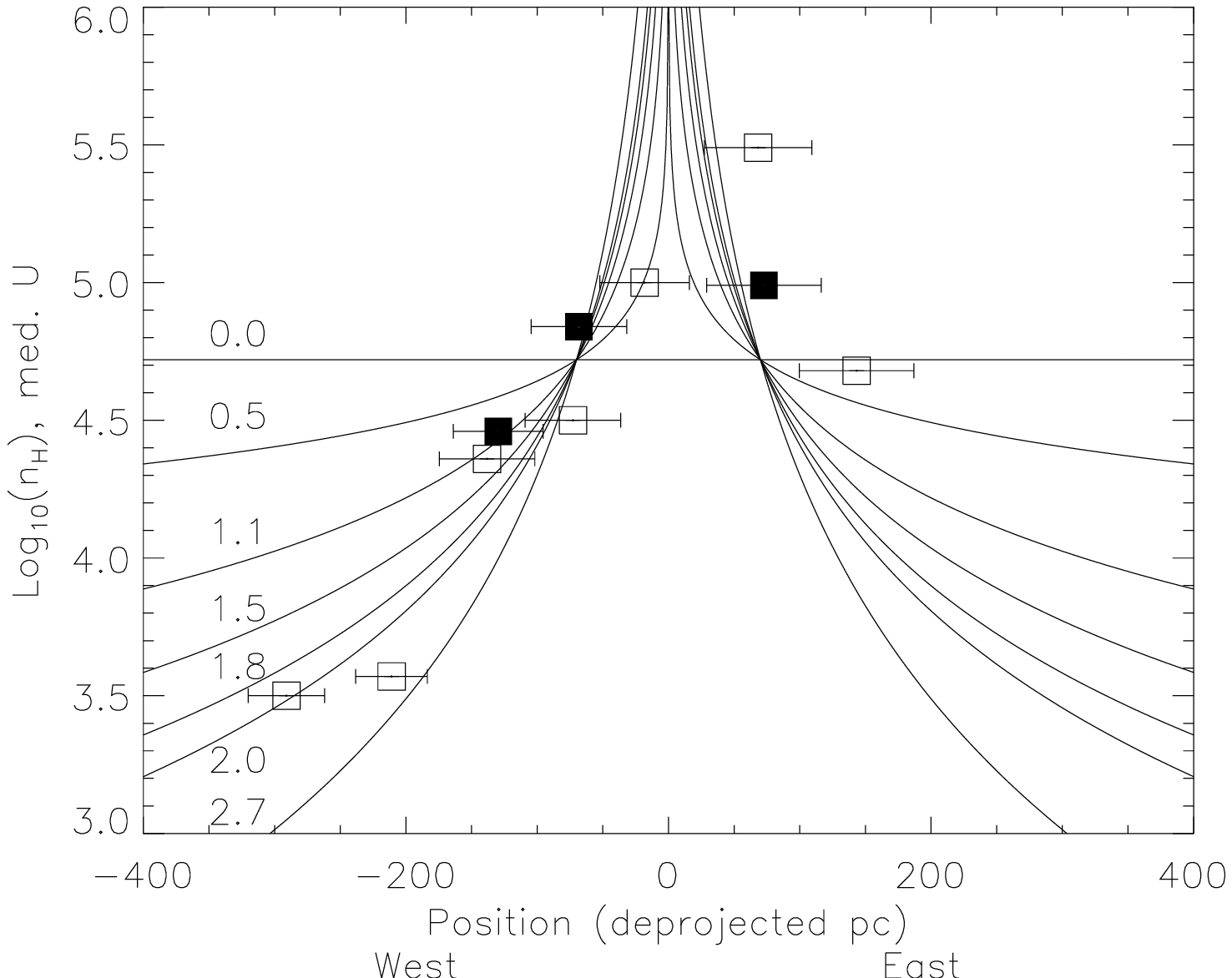]{Logarithm of the hydrogen volume number density	 
for medium ionization state model clouds as a function of radial distance from the AGN.	 
See the Figure~5 caption for a description of the plot symbols.}

\figcaption[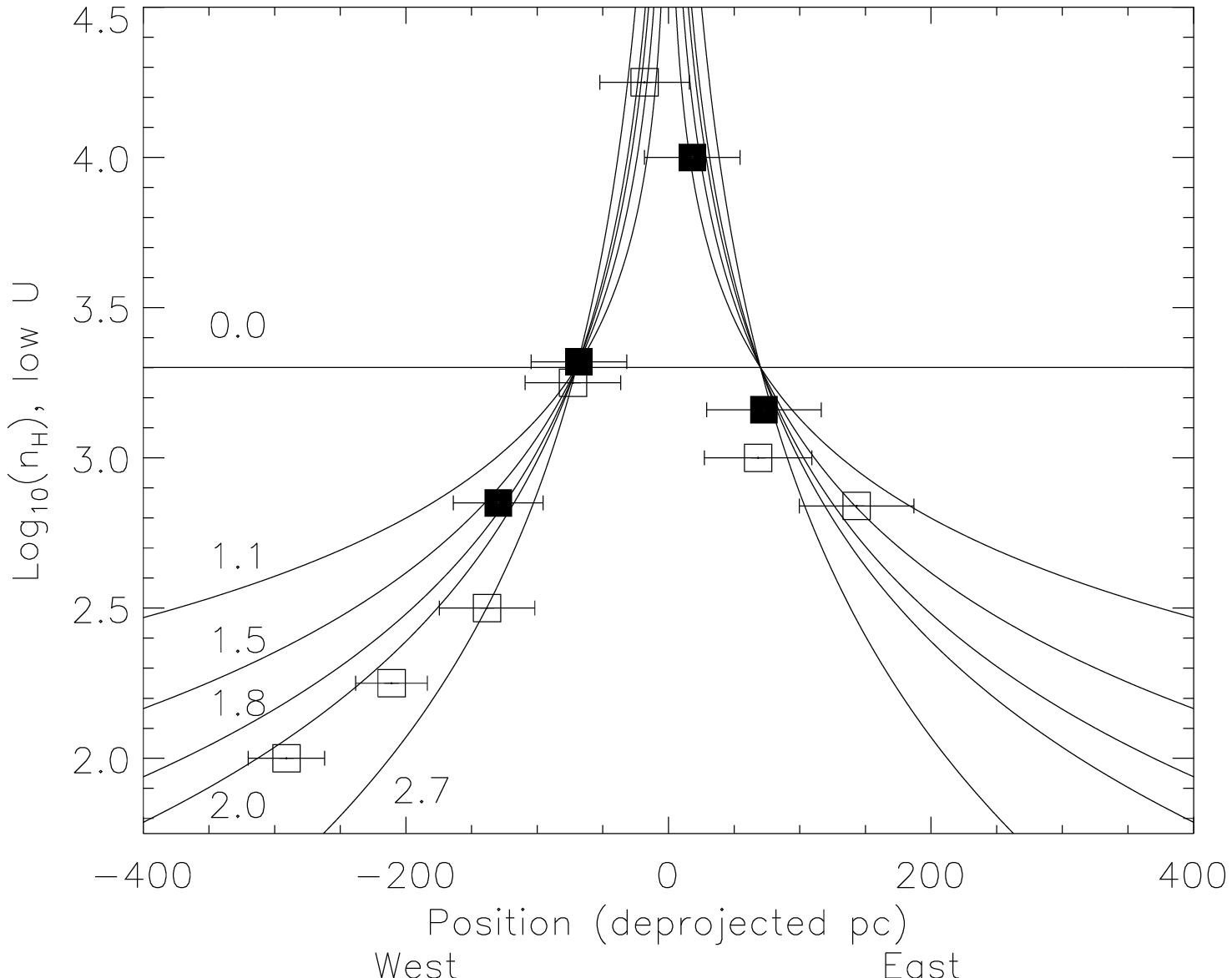]{Logarithm of the hydrogen volume number density	 
for low ionization state model clouds as a function of radial distance from the AGN.	 
See the Figure~5 caption for a description of the plot symbols.}

\begin{deluxetable}{llccccccc}
\label{table1}
\tabletypesize{\scriptsize}
%\rotate
\tablewidth{0pc}
\tablenum{1}              %  model_input_table.tex
\tablecolumns{9}
\tablecaption{CLOUDY Photoionization Model Input Parameters}
\tablehead{
\colhead{Bin} &
\colhead{Model} &
\colhead{Distance} &
\colhead{Screen} &
\colhead{Screen} &
\colhead{Fractional} &
\colhead{n$_{H}$} &
\colhead{N$_{c}$} \\
\colhead{} &
\colhead{Component} &
\colhead{} &
\colhead{U} &
\colhead{N$_{c}$} & 
\colhead{Contribution} &
\colhead{} & 
\colhead{}\\
\colhead{} &
\colhead{} &
\colhead{(pc)} &
\colhead{} &
\colhead{(cm$^{-2}$)} &
\colhead{} &
\colhead{(cm$^{-3}$)} &
\colhead{(cm$^{-2}$)}\\
\colhead{(1)} &
\colhead{(2)} &
\colhead{(3)} &
\colhead{(4)} &
\colhead{(5)} &
\colhead{(6)} &
\colhead{(7)} &
\colhead{(8)}
}
\startdata
( 0$\farcs$5E, r, 0$\farcs$3) & high & 143 & 10$^{ -1.5}$ & 10$^{ 20.4}$ & 0.35 & 10$^{ 2.74}$ & 10$^{ 22.5}$ \\
                             & med. & 143 & 10$^{ -1.5}$ & 10$^{ 20.4}$ & 0.35 & 10$^{ 4.68}$ & 10$^{ 20.0}$ \\
                             & low  & 155 & 10$^{ -2.5}$ & 10$^{ 20.7}$ & 0.30 & 10$^{ 2.84}$ & 10$^{ 22.5}$ \\
( 0$\farcs$3E, b, 0$\farcs$3) & high &  73 & 10$^{ -1.5}$ & 10$^{ 20.4}$ & 0.25 & 10$^{ 3.49}$ & 10$^{ 22.0}$ \\
                             & med. &  73 & 10$^{ -1.5}$ & 10$^{ 20.4}$ & 0.45 & 10$^{ 4.99}$ & 10$^{ 21.1}$ \\
                             & low  &  82 & 10$^{ -3.0}$ & 10$^{ 21.9}$ & 0.30 & 10$^{ 3.16}$ & 10$^{ 22.0}$ \\
( 0$\farcs$3E, r, 0$\farcs$3) & high &  68 & 10$^{ -1.5}$ & 10$^{ 20.4}$ & 0.25 & 10$^{ 3.99}$ & 10$^{ 21.5}$ \\
                             & med. &  68 & 10$^{ -1.5}$ & 10$^{ 20.4}$ & 0.45 & 10$^{ 5.49}$ & 10$^{ 20.7}$ \\
\enddata                                                                                                        

\end{deluxetable}

%%%%%%%%%%%%%%%%%%%%%%%%%%%%%%%%%%%%%%%%%%%%%%%%%%%%%%%%%%%%%%%%%%%%%%%%%%%%%%

\begin{deluxetable}{lrcrrrrrr}
\label{table2}
%\rotate
\tablecolumns{9}

\tabletypesize{\scriptsize}
\tablewidth{0pc}
\tablenum{2}
\tablecaption{Line Ratios (Relative to H$\beta$) from 
Model Components, Composite and Observations for 
Position (0$\farcs$5E,~r,~0$\farcs$3)}
\tablehead{
\colhead{} &
\colhead{} &
\colhead{Line} &
\colhead{High-U} &
\colhead{Med.-U} &
\colhead{Low-U} &
\colhead{Composite} &
\colhead{Observed} &
\colhead{Model/Data}\\

\colhead{} &
\colhead{} &
\colhead{Match\tablenotemark{a}} &
\colhead{Model\tablenotemark{b}} &
\colhead{Model\tablenotemark{c}} &
\colhead{Model\tablenotemark{d}} &
\colhead{Model} &
\colhead{Data\tablenotemark{e}} &
\colhead{Ratio}\\

\colhead{(1)} &
\colhead{(2)} &
\colhead{(3)} &
\colhead{(4)} &
\colhead{(5)} &
\colhead{(6)} &
\colhead{(7)} &
\colhead{(8)} &
\colhead{(9)}
}
\startdata

Ly$\alpha$  & 1216 &    0  &    24.05  & 23.78   &   5.92  & 18.52   &  38.97~$\pm$~38.35 &    0.5~$\pm$~0.5 \\
NV          & 1240 &    1  &     0.71  &  0.06   &   0.11  &  0.30   &   0.56~$\pm$~0.84  &    0.5~$\pm$~0.8 \\
CIV         & 1549 &    1  &     4.02  &  0.88   &   0.74  &  1.93   &   1.82~$\pm$~1.08  &    1.1~$\pm$~0.7 \\
CIII$]$     & 1909 &    0  &     0.12  &  1.88   &   1.59  &  1.18   &   3.43~$\pm$~2.54  &    0.3~$\pm$~0.2 \\
CII$]$      & 2326 &    0  &     0.00  &  0.10   &   1.90  &  0.60   &   1.84~$\pm$~1.17  &    0.3~$\pm$~0.2 \\
$[$NeIV$]$  & 2424 &    1  &     0.15  &  0.15   &   0.02  &  0.11   &   0.16~$\pm$~0.14  &    0.7~$\pm$~0.6 \\
MgII        & 2798 &    1  &     0.01  &  0.24   &   2.45  &  0.82   &   1.03~$\pm$~0.42  &    0.8~$\pm$~0.3 \\
$[$NeV$]$   & 3426 &    1  &     0.49  &  0.03   &   0.01  &  0.19   &   0.36~$\pm$~0.17  &    0.5~$\pm$~0.2 \\
\enddata

\tablenotetext{a}{1=Model and data line ratios relative to H$\beta$ match 
                   within a factor of 2; 0=no match.}
\tablenotetext{b}{High-U model. Boundary:  m; 
                  log$_{10}$(U)= -0.29; 
                  log$_{10}$(n$_{H}$)= 2.74~cm$^{-3}$;  
                  cloud log$_{10}$(N$_{c}$)= 22.5~cm$^{-2}$;  
                  fractional contribution = 0.35.}
\tablenotetext{c}{Medium-U model. Boundary:  m; 
                  log$_{10}$(U)= -2.23; 
                  log$_{10}$(n$_{H}$)= 4.68~cm$^{-3}$;  
                  cloud log$_{10}$(N$_{c}$)= 20.0~cm$^{-2}$;  
                  fractional contribution = 0.35.}
\tablenotetext{d}{~Low-U model. Boundary:  m; 
                  log$_{10}$(U)= -2.69; log$_{10}$(n$_{H}$)= 2.84~cm$^{-3}$;  
                  cloud log$_{10}$(N$_{c}$)= 22.5~cm$^{-2}$;  
                  fractional contribution = 0.30;
                  screen log$_{10}$(N$_{c}$)= 20.7~cm$^{-2}$.}
\tablenotetext{e}{We corrected all observed fluxes for Galactic extinction using the \citet{sav79} 
                extinction curve with E(B-V)~=~0.19.  We applied an additional correction for   
                Mrk~3 internal extinction using the LMC curve of \citet{koo81}.
                For this measurement bin the internal extinction correction 
                corresponded to E(B-V)~=~0.31.
                See Paper~I for details.}
\tablenotetext{f}{~Logarithm of model H$\beta$ flux.  Units: log$_{10}$(ergs~s$^{-1}$~cm$^{-2}$).}
\tablenotetext{g}{Observed H$\beta$ flux. Units: 10$^{-15}$~ergs~s$^{-1}$~cm$^{-2}$.}

\end{deluxetable}

\newpage

\begin{deluxetable}{lrcrrrrrr}
\label{table3}
%\rotate
\tablecolumns{9}

\tabletypesize{\scriptsize}
\tablewidth{0pc}
\tablenum{3}
\tablecaption{Line Ratios (Relative to H$\beta$) from 
Model Components, Composite and Observations for 
Position (0$\farcs$3E,~b,~0$\farcs$3)}
\tablehead{
\colhead{} &
\colhead{} &
\colhead{Line} &
\colhead{High-U} &
\colhead{Med.-U} &
\colhead{Low-U} &
\colhead{Composite} &
\colhead{Observed} &
\colhead{Model/Data}\\

\colhead{} &
\colhead{} &
\colhead{Match\tablenotemark{a}} &
\colhead{Model\tablenotemark{b}} &
\colhead{Model\tablenotemark{c}} &
\colhead{Model\tablenotemark{d}} &
\colhead{Model} &
\colhead{Data\tablenotemark{e}} &
\colhead{Ratio}\\

\colhead{(1)} &
\colhead{(2)} &
\colhead{(3)} &
\colhead{(4)} &
\colhead{(5)} &
\colhead{(6)} &
\colhead{(7)} &
\colhead{(8)} &
\colhead{(9)}
}
\startdata

Ly$\alpha$  & 1216 &   1   & 26.07   &  4.13  &   2.65  &   9.17  &   9.90~$\pm$~7.97   &   0.9~$\pm$~0.7 \\
NV          & 1240 &   1   &  1.69   &  0.02  &   0.00  &   0.43  &   0.46~$\pm$~0.76   &   0.9~$\pm$~1.5 \\
CIV         & 1549 &   0   & 11.44   &  0.61  &   0.55  &   3.30  &   0.57~$\pm$~0.34   &   5.7~$\pm$~3.4 \\
CIII$]$     & 1909 &   1   &  0.32   &  1.64  &   0.03  &   0.83  &   1.65~$\pm$~0.97   &   0.5~$\pm$~0.3 \\
CII$]$      & 2326 &   1   &  0.00   &  0.21  &   1.09  &   0.42  &   0.68~$\pm$~0.36   &   0.6~$\pm$~0.3 \\
$[$NeIV$]$  & 2424 &   1   &  0.48   &  0.08  &   0.00  &   0.16  &   0.31~$\pm$~0.15   &   0.5~$\pm$~0.2 \\
MgII        & 2798 &   1   &  0.02   &  0.61  &   2.88  &   1.14  &   0.73~$\pm$~0.24   &   1.6~$\pm$~0.5 \\
$[$NeV$]$   & 3426 &   1   &  1.40   &  0.03  &   0.00  &   0.37  &   0.63~$\pm$~0.24   &   0.6~$\pm$~0.2 \\
\enddata

\tablenotetext{a}{1=Model and data line ratios relative to H$\beta$ match 
                   within a factor of 2; 0=no match.}
\tablenotetext{b}{High-U model. Boundary:  m; 
                  log$_{10}$(U)= -0.45; 
                  log$_{10}$(n$_{H}$)= 3.49~cm$^{-3}$;  
                  cloud log$_{10}$(N$_{c}$)= 22.0~cm$^{-2}$;  
                  fractional contribution = 0.25.}
\tablenotetext{c}{Medium-U model. Boundary:  r; 
                  log$_{10}$(U)= -1.95; 
                  log$_{10}$(n$_{H}$)= 4.99~cm$^{-3}$;  
                  cloud log$_{10}$(N$_{c}$)= 21.1~cm$^{-2}$;  
                  fractional contribution = 0.45.}
\tablenotetext{d}{~Low-U model. Boundary:  m; 
                  log$_{10}$(U)= -3.08; log$_{10}$(n$_{H}$)= 3.16~cm$^{-3}$;  
                  cloud log$_{10}$(N$_{c}$)= 22.0~cm$^{-2}$;  
                  fractional contribution = 0.30;
                  screen log$_{10}$(N$_{c}$)= 21.9~cm$^{-2}$.}
\tablenotetext{e}{We corrected all observed fluxes for Galactic extinction using the \citet{sav79} 
                extinction curve with E(B-V)~=~0.19.  We applied an additional correction for   
                Mrk~3 internal extinction using the LMC curve of \citet{koo81}.
                For this measurement bin the internal extinction correction 
                corresponded to E(B-V)~=~0.26.
                See Paper~I for details.}
\tablenotetext{f}{~Logarithm of model H$\beta$ flux.  Units: log$_{10}$(ergs~s$^{-1}$~cm$^{-2}$).}
\tablenotetext{g}{Observed H$\beta$ flux. Units: 10$^{-15}$~ergs~s$^{-1}$~cm$^{-2}$.}

\end{deluxetable}

\newpage

\begin{deluxetable}{lrcrrrrrr}
\label{table4}
%\rotate
\tablecolumns{9}

\tabletypesize{\scriptsize}
\tablewidth{0pc}
\tablenum{4}
\tablecaption{Line Ratios (Relative to H$\beta$) from 
Model Components, Composite and Observations for 
Position (0$\farcs$3E,~r,~0$\farcs$3)}
\tablehead{
\colhead{} &
\colhead{} &
\colhead{Line} &
\colhead{High-U} &
\colhead{Med.-U} &
\colhead{Low-U} &
\colhead{Composite} &
\colhead{Observed} &
\colhead{Model/Data}\\

\colhead{} &
\colhead{} &
\colhead{Match\tablenotemark{a}} &
\colhead{Model\tablenotemark{b}} &
\colhead{Model\tablenotemark{c}} &
\colhead{Model\tablenotemark{d}} &
\colhead{Model} &
\colhead{Data\tablenotemark{e}} &
\colhead{Ratio}\\

\colhead{(1)} &
\colhead{(2)} &
\colhead{(3)} &
\colhead{(4)} &
\colhead{(5)} &
\colhead{(6)} &
\colhead{(7)} &
\colhead{(8)} &
\colhead{(9)}
}
\startdata

Ly$\alpha$  & 1216 & 1   &   28.43  &   8.17  &   2.76  &  11.61  &  19.97~$\pm$~11.55 &   0.6~$\pm$~0.3 \\
NV          & 1240 & 1   &    0.89  &   0.01  &   0.01  &   0.23  &   0.22~$\pm$~0.13  &   1.0~$\pm$~0.6 \\
CIV         & 1549 & 1   &   11.27  &   0.21  &   0.60  &   3.09  &   2.11~$\pm$~0.74  &   1.5~$\pm$~0.5 \\
CIII$]$     & 1909 & 1   &    0.88  &   1.44  &   0.09  &   0.89  &   0.77~$\pm$~0.32  &   1.2~$\pm$~0.5 \\
CII$]$      & 2326 & 1   &    0.00  &   0.37  &   1.77  &   0.69  &   0.81~$\pm$~0.31  &   0.9~$\pm$~0.3 \\
$[$NeIV$]$  & 2424 & 0   &    0.63  &   0.01  &   0.00  &   0.16  &   0.38~$\pm$~0.13  &   0.4~$\pm$~0.1 \\
MgII        & 2798 & 1   &    0.04  &   1.03  &   3.29  &   1.46  &   0.96~$\pm$~0.23  &   1.5~$\pm$~0.4 \\
$[$NeV$]$   & 3426 & 1   &    1.24  &   0.00  &   0.00  &   0.31  &   0.41~$\pm$~0.12  &   0.8~$\pm$~0.2 \\
\enddata

\tablenotetext{a}{1=Model and data line ratios relative to H$\beta$ match 
                   within a factor of 2; 0=no match.}
\tablenotetext{b}{High-U model. Boundary:  m; 
                  log$_{10}$(U)= -0.89; 
                  log$_{10}$(n$_{H}$)= 3.99~cm$^{-3}$;  
                  cloud log$_{10}$(N$_{c}$)= 21.5~cm$^{-2}$;  
                  fractional contribution = 0.25.}
\tablenotetext{c}{Medium-U model. Boundary:  r; 
                  log$_{10}$(U)= -2.39; 
                  log$_{10}$(n$_{H}$)= 5.49~cm$^{-3}$;  
                  cloud log$_{10}$(N$_{c}$)= 20.7~cm$^{-2}$;  
                  fractional contribution = 0.45.}
\tablenotetext{d}{~Low-U model. Boundary:  m; 
                  log$_{10}$(U)= -2.83; log$_{10}$(n$_{H}$)= 3.00~cm$^{-3}$;  
                  cloud log$_{10}$(N$_{c}$)= 22.0~cm$^{-2}$;  
                  fractional contribution = 0.30;
                  screen log$_{10}$(N$_{c}$)= 21.9~cm$^{-2}$.}
\tablenotetext{e}{We corrected all observed fluxes for Galactic extinction using the \citet{sav79} 
                extinction curve with E(B-V)~=~0.19.  We applied an additional correction for   
                Mrk~3 internal extinction using the LMC curve of \citet{koo81}.
                For this measurement bin the internal extinction correction 
                corresponded to E(B-V)~=~0.24.
                See Paper~I for details.}
\tablenotetext{f}{~Logarithm of model H$\beta$ flux.  Units: log$_{10}$(ergs~s$^{-1}$~cm$^{-2}$).}
\tablenotetext{g}{Observed H$\beta$ flux. Units: 10$^{-15}$~ergs~s$^{-1}$~cm$^{-2}$.}

\end{deluxetable}

\newpage

\begin{deluxetable}{lrcrrrrrr}
\label{table5}
%\rotate
\tablecolumns{9}

\tabletypesize{\scriptsize}
\tablewidth{0pc}
\tablenum{5}
\tablecaption{Line Ratios (Relative to H$\beta$) from 
Model Components, Composite and Observations for 
Position (0$\farcs$0E,~b,~0$\farcs$3)}
\tablehead{
\colhead{} &
\colhead{} &
\colhead{Line} &
\colhead{High-U} &
\colhead{Med.-U} &
\colhead{Low-U} &
\colhead{Composite} &
\colhead{Observed} &
\colhead{Model/Data}\\

\colhead{} &
\colhead{} &
\colhead{Match\tablenotemark{a}} &
\colhead{Model\tablenotemark{b}} &
\colhead{Model\tablenotemark{c}} &
\colhead{Model\tablenotemark{d}} &
\colhead{Model} &
\colhead{Data\tablenotemark{e}} &
\colhead{Ratio}\\

\colhead{(1)} &
\colhead{(2)} &
\colhead{(3)} &
\colhead{(4)} &
\colhead{(5)} &
\colhead{(6)} &
\colhead{(7)} &
\colhead{(8)} &
\colhead{(9)}
}
\startdata

Ly$\alpha$  & 1216 &  0  &     36.59  &  \nodata   &    4.39  &   26.93  &    6.78~$\pm$~4.69  &     4.0~$\pm$~2.8  \\
NV          & 1240 &  1  &      0.67  &  \nodata   &    0.14  &    0.51  &    0.40~$\pm$~0.28  &     1.3~$\pm$~0.9  \\
CIV         & 1549 &  0  &      7.30  &  \nodata   &    1.79  &    5.65  &    0.41~$\pm$~0.22  &    13.7~$\pm$~7.4  \\
CIII$]$     & 1909 &  1  &      0.50  &  \nodata   &    0.79  &    0.59  &    0.71~$\pm$~0.39  &     0.8~$\pm$~0.4  \\
CII$]$      & 2326 &  1  &      0.00  &  \nodata   &    2.85  &    0.85  &    0.60~$\pm$~0.27  &     1.4~$\pm$~0.6  \\
$[$NeIV$]$  & 2424 &  1  &      0.21  &  \nodata   &    0.02  &    0.15  &    0.22~$\pm$~0.09  &     0.7~$\pm$~0.3  \\
MgII        & 2798 &  0  &      0.03  &  \nodata   &    3.81  &    1.16  &    0.40~$\pm$~0.11  &     2.9~$\pm$~0.8  \\
$[$NeV$]$   & 3426 &  1  &      0.69  &  \nodata   &    0.02  &    0.49  &    0.44~$\pm$~0.16  &     1.1~$\pm$~0.4  \\
\enddata

\tablenotetext{a}{1=Model and data line ratios relative to H$\beta$ match 
                   within a factor of 2; 0=no match.}
\tablenotetext{b}{High-U model. Boundary:  m; 
                  log$_{10}$(U)= -0.75; 
                  log$_{10}$(n$_{H}$)= 5.00~cm$^{-3}$;  
                  cloud log$_{10}$(N$_{c}$)= 22.0~cm$^{-2}$;  
                  fractional contribution = 0.70.}
\tablenotetext{c}{No medium-U model component required for this observed kinematic component.}
\tablenotetext{d}{~Low-U model. Boundary:  m; 
                  log$_{10}$(U)= -2.52; log$_{10}$(n$_{H}$)= 4.00~cm$^{-3}$;  
                  cloud log$_{10}$(N$_{c}$)= 23.0~cm$^{-2}$;  
                  fractional contribution = 0.30;
                  screen log$_{10}$(N$_{c}$)= 21.6~cm$^{-2}$.}
\tablenotetext{e}{We corrected all observed fluxes for Galactic extinction using the \citet{sav79} 
                extinction curve with E(B-V)~=~0.19.  We applied an additional correction for   
                Mrk~3 internal extinction using the LMC curve of \citet{koo81}.
                For this measurement bin the internal extinction correction 
                corresponded to E(B-V)~=~0.22.
                See Paper~I for details.}
\tablenotetext{f}{~Logarithm of model H$\beta$ flux.  Units: log$_{10}$(ergs~s$^{-1}$~cm$^{-2}$).}
\tablenotetext{g}{Observed H$\beta$ flux. Units: 10$^{-15}$~ergs~s$^{-1}$~cm$^{-2}$.}

\end{deluxetable}

\newpage

\begin{deluxetable}{lrcrrrrrr}
\label{table6}
%\rotate
\tablecolumns{9}

\tabletypesize{\scriptsize}
\tablewidth{0pc}
\tablenum{6}
\tablecaption{Line Ratios (Relative to H$\beta$) from 
Model Components, Composite and Observations for 
Position (0$\farcs$0W,~r,~0$\farcs$3)}
\tablehead{
\colhead{} &
\colhead{} &
\colhead{Line} &
\colhead{High-U} &
\colhead{Med.-U} &
\colhead{Low-U} &
\colhead{Composite} &
\colhead{Observed} &
\colhead{Model/Data}\\

\colhead{} &
\colhead{} &
\colhead{Match\tablenotemark{a}} &
\colhead{Model\tablenotemark{b}} &
\colhead{Model\tablenotemark{c}} &
\colhead{Model\tablenotemark{d}} &
\colhead{Model} &
\colhead{Data\tablenotemark{e}} &
\colhead{Ratio}\\

\colhead{(1)} &
\colhead{(2)} &
\colhead{(3)} &
\colhead{(4)} &
\colhead{(5)} &
\colhead{(6)} &
\colhead{(7)} &
\colhead{(8)} &
\colhead{(9)}
}
\startdata

Ly$\alpha$  & 1216 &  0  &    34.80  &   1.37  &   2.19  &   8.30  &  54.35~$\pm$~14.72  &   0.2~$\pm$~0.1  \\
NV          & 1240 &  1  &     5.10  &   1.05  &   0.06  &   1.56  &   2.46~$\pm$~0.66   &   0.6~$\pm$~0.2  \\
CIV         & 1549 &  1  &    37.32  &   5.63  &   0.91  &  10.55  &   6.51~$\pm$~1.12   &   1.6~$\pm$~0.3  \\
CIII$]$     & 1909 &  0  &     1.12  &   2.68  &   0.17  &   1.62  &   4.99~$\pm$~1.07   &   0.3~$\pm$~0.1  \\
CII$]$      & 2326 &  1  &     0.00  &   0.46  &   1.23  &   0.60  &   1.13~$\pm$~0.26   &   0.5~$\pm$~0.1  \\
$[$NeIV$]$  & 2424 &  1  &     1.26  &   0.72  &   0.00  &   0.61  &   1.17~$\pm$~0.22   &   0.5~$\pm$~0.1  \\
MgII        & 2798 &  1  &     0.05  &   0.43  &   2.28  &   0.91  &   0.70~$\pm$~0.12   &   1.3~$\pm$~0.2  \\
$[$NeV$]$   & 3426 &  1  &     4.31  &   1.93  &   0.00  &   1.83  &   1.43~$\pm$~0.24   &   1.3~$\pm$~0.2  \\
\enddata

\tablenotetext{a}{1=Model and data line ratios relative to H$\beta$ match 
                   within a factor of 2; 0=no match.}
\tablenotetext{b}{High-U model. Boundary:  m; 
                  log$_{10}$(U)= -0.60; 
                  log$_{10}$(n$_{H}$)= 4.85~cm$^{-3}$;  
                  cloud log$_{10}$(N$_{c}$)= 21.5~cm$^{-2}$;  
                  fractional contribution = 0.20.}
\tablenotetext{c}{Medium-U model. Boundary:  r; 
                  log$_{10}$(U)= -0.75; 
                  log$_{10}$(n$_{H}$)= 5.00~cm$^{-3}$;  
                  cloud log$_{10}$(N$_{c}$)= 22.3~cm$^{-2}$;  
                  fractional contribution = 0.50.}
\tablenotetext{d}{~Low-U model. Boundary:  m; 
                  log$_{10}$(U)= -2.77; log$_{10}$(n$_{H}$)= 4.25~cm$^{-3}$;  
                  cloud log$_{10}$(N$_{c}$)= 23.5~cm$^{-2}$;  
                  fractional contribution = 0.30;
                  screen log$_{10}$(N$_{c}$)= 21.6~cm$^{-2}$.}
\tablenotetext{e}{We corrected all observed fluxes for Galactic extinction using the \citet{sav79} 
                extinction curve with E(B-V)~=~0.19.  We applied an additional correction for   
                Mrk~3 internal extinction using the LMC curve of \citet{koo81}.
                For this measurement bin the internal extinction correction 
                corresponded to E(B-V)~=~0.40.
                See Paper~I for details.}
\tablenotetext{f}{~Logarithm of model H$\beta$ flux.  Units: log$_{10}$(ergs~s$^{-1}$~cm$^{-2}$).}
\tablenotetext{g}{Observed H$\beta$ flux. Units: 10$^{-15}$~ergs~s$^{-1}$~cm$^{-2}$.}

\end{deluxetable}

\newpage

\begin{deluxetable}{lrcrrrrrr}
\label{table7}
%\rotate
\tablecolumns{9}

\tabletypesize{\scriptsize}
\tablewidth{0pc}
\tablenum{7}
\tablecaption{Line Ratios (Relative to H$\beta$) from 
Model Components, Composite and Observations for 
Position (-0$\farcs$3W,~b,~0$\farcs$3)}
\tablehead{
\colhead{} &
\colhead{} &
\colhead{Line} &
\colhead{High-U} &
\colhead{Med.-U} &
\colhead{Low-U} &
\colhead{Composite} &
\colhead{Observed} &
\colhead{Model/Data}\\

\colhead{} &
\colhead{} &
\colhead{Match\tablenotemark{a}} &
\colhead{Model\tablenotemark{b}} &
\colhead{Model\tablenotemark{c}} &
\colhead{Model\tablenotemark{d}} &
\colhead{Model} &
\colhead{Data\tablenotemark{e}} &
\colhead{Ratio}\\

\colhead{(1)} &
\colhead{(2)} &
\colhead{(3)} &
\colhead{(4)} &
\colhead{(5)} &
\colhead{(6)} &
\colhead{(7)} &
\colhead{(8)} &
\colhead{(9)}
}
\startdata

Ly$\alpha$  & 1216 &  1  &    28.14  &  27.01  &   6.60  &  21.34  &  18.58~$\pm$~5.76   &     1.1~$\pm$~0.3   \\
NV          & 1240 &  1  &     0.49  &   0.53  &   0.03  &   0.36  &   0.69~$\pm$~0.22   &     0.5~$\pm$~0.2  \\
CIV         & 1549 &  1  &     5.08  &  10.96  &   2.14  &   5.96  &   3.06~$\pm$~0.60   &     1.9~$\pm$~0.4  \\
CIII$]$     & 1909 &  1  &     0.35  &   5.13  &   0.14  &   1.72  &   1.36~$\pm$~0.37   &     1.3~$\pm$~0.4  \\
CII$]$      & 2326 &  1  &     0.00  &   0.06  &   1.15  &   0.36  &   0.64~$\pm$~0.13   &     0.6~$\pm$~0.1  \\
$[$NeIV$]$  & 2424 &  1  &     0.25  &   1.53  &   0.00  &   0.56  &   0.46~$\pm$~0.09   &     1.2~$\pm$~0.2  \\
MgII        & 2798 &  1  &     0.02  &   0.12  &   2.36  &   0.75  &   0.75~$\pm$~0.10   &     1.0~$\pm$~0.1  \\
$[$NeV$]$   & 3426 &  1  &     0.57  &   0.99  &   0.00  &   0.52  &   0.78~$\pm$~0.12   &     0.7~$\pm$~0.1  \\
\enddata

\tablenotetext{a}{1=Model and data line ratios relative to H$\beta$ match 
                   within a factor of 2; 0=no match.}
\tablenotetext{b}{High-U model. Boundary:  m; 
                  log$_{10}$(U)= -0.74; 
                  log$_{10}$(n$_{H}$)= 3.84~cm$^{-3}$;  
                  cloud log$_{10}$(N$_{c}$)= 22.0~cm$^{-2}$;  
                  fractional contribution = 0.40.}
\tablenotetext{c}{Medium-U model. Boundary:  m; 
                  log$_{10}$(U)= -1.74; 
                  log$_{10}$(n$_{H}$)= 4.84~cm$^{-3}$;  
                  cloud log$_{10}$(N$_{c}$)= 20.0~cm$^{-2}$;  
                  fractional contribution = 0.30.}
\tablenotetext{d}{~Low-U model. Boundary:  m; 
                  log$_{10}$(U)= -2.95; log$_{10}$(n$_{H}$)= 3.32~cm$^{-3}$;  
                  cloud log$_{10}$(N$_{c}$)= 22.5~cm$^{-2}$;  
                  fractional contribution = 0.30;
                  screen log$_{10}$(N$_{c}$)= 21.6~cm$^{-2}$.}
\tablenotetext{e}{We corrected all observed fluxes for Galactic extinction using the \citet{sav79} 
                extinction curve with E(B-V)~=~0.19.  We applied an additional correction for   
                Mrk~3 internal extinction using the LMC curve of \citet{koo81}.
                For this measurement bin the internal extinction correction 
                corresponded to E(B-V)~=~0.22.
                See Paper~I for details.}
\tablenotetext{f}{~Logarithm of model H$\beta$ flux.  Units: log$_{10}$(ergs~s$^{-1}$~cm$^{-2}$).}
\tablenotetext{g}{Observed H$\beta$ flux. Units: 10$^{-15}$~ergs~s$^{-1}$~cm$^{-2}$.}

\end{deluxetable}

\newpage

\begin{deluxetable}{lrcrrrrrr}
\label{table8}
%\rotate
\tablecolumns{9}

\tabletypesize{\scriptsize}
\tablewidth{0pc}
\tablenum{8}
\tablecaption{Line Ratios (Relative to H$\beta$) from 
Model Components, Composite and Observations for 
Position (-0$\farcs$3W,~r,~0$\farcs$3)}
\tablehead{
\colhead{} &
\colhead{} &
\colhead{Line} &
\colhead{High-U} &
\colhead{Med.-U} &
\colhead{Low-U} &
\colhead{Composite} &
\colhead{Observed} &
\colhead{Model/Data}\\

\colhead{} &
\colhead{} &
\colhead{Match\tablenotemark{a}} &
\colhead{Model\tablenotemark{b}} &
\colhead{Model\tablenotemark{c}} &
\colhead{Model\tablenotemark{d}} &
\colhead{Model} &
\colhead{Data\tablenotemark{e}} &
\colhead{Ratio}\\

\colhead{(1)} &
\colhead{(2)} &
\colhead{(3)} &
\colhead{(4)} &
\colhead{(5)} &
\colhead{(6)} &
\colhead{(7)} &
\colhead{(8)} &
\colhead{(9)}
}
\startdata

Ly$\alpha$  & 1216 &   1  &  27.23 & 16.97   &   10.88 & 19.25   &   29.68~$\pm$~7.08 &     0.6~$\pm$~0.1  \\
NV          & 1240 &   0  &   0.46 &  0.86   &    0.04 &  0.45   &    1.29~$\pm$~0.31 &     0.3~$\pm$~0.1  \\
CIV         & 1549 &   1  &   3.30 & 13.27   &    3.34 &  6.30   &    6.28~$\pm$~0.90 &     1.0~$\pm$~0.1  \\
CIII$]$     & 1909 &   1  &   0.67 &  4.51   &    0.19 &  1.68   &    2.20~$\pm$~0.52 &     0.8~$\pm$~0.2  \\
CII$]$      & 2326 &   1  &   0.00 &  0.03   &    1.49 &  0.46   &    0.59~$\pm$~0.09 &     0.8~$\pm$~0.1  \\
$[$NeIV$]$  & 2424 &   1  &   0.13 &  2.08   &    0.00 &  0.68   &    0.76~$\pm$~0.11 &     0.9~$\pm$~0.1  \\
MgII        & 2798 &   1  &   0.04 &  0.08   &    2.96 &  0.93   &    1.05~$\pm$~0.10 &     0.9~$\pm$~0.1  \\
$[$NeV$]$   & 3426 &   1  &   0.39 &  1.82   &    0.00 &  0.70   &    1.07~$\pm$~0.12 &     0.7~$\pm$~0.1  \\
\enddata

\tablenotetext{a}{1=Model and data line ratios relative to H$\beta$ match 
                   within a factor of 2; 0=no match.}
\tablenotetext{b}{High-U model. Boundary:  m; 
                  log$_{10}$(U)= -0.46; 
                  log$_{10}$(n$_{H}$)= 3.50~cm$^{-3}$;  
                  cloud log$_{10}$(N$_{c}$)= 22.5~cm$^{-2}$;  
                  fractional contribution = 0.40.}
\tablenotetext{c}{Medium-U model. Boundary:  m; 
                  log$_{10}$(U)= -1.46; 
                  log$_{10}$(n$_{H}$)= 4.50~cm$^{-3}$;  
                  cloud log$_{10}$(N$_{c}$)= 20.5~cm$^{-2}$;  
                  fractional contribution = 0.30.}
\tablenotetext{d}{~Low-U model. Boundary:  m; 
                  log$_{10}$(U)= -2.98; log$_{10}$(n$_{H}$)= 3.25~cm$^{-3}$;  
                  cloud log$_{10}$(N$_{c}$)= 22.0~cm$^{-2}$;  
                  fractional contribution = 0.30;
                  screen log$_{10}$(N$_{c}$)= 21.6~cm$^{-2}$.}
\tablenotetext{e}{We corrected all observed fluxes for Galactic extinction using the \citet{sav79} 
                extinction curve with E(B-V)~=~0.19.  We applied an additional correction for   
                Mrk~3 internal extinction using the LMC curve of \citet{koo81}.
                For this measurement bin the internal extinction correction 
                corresponded to E(B-V)~=~0.21.
                See Paper~I for details.}
\tablenotetext{f}{~Logarithm of model H$\beta$ flux.  Units: log$_{10}$(ergs~s$^{-1}$~cm$^{-2}$).}
\tablenotetext{g}{Observed H$\beta$ flux. Units: 10$^{-15}$~ergs~s$^{-1}$~cm$^{-2}$.}

\end{deluxetable}

\newpage

\begin{deluxetable}{lrcrrrrrr}
\label{table9}
%\rotate
\tablecolumns{9}

\tabletypesize{\scriptsize}
\tablewidth{0pc}
\tablenum{9}
\tablecaption{Line Ratios (Relative to H$\beta$) from 
Model Components, Composite and Observations for 
Position (-0$\farcs$5W,~b,~0$\farcs$2)}
\tablehead{
\colhead{} &
\colhead{} &
\colhead{Line} &
\colhead{High-U} &
\colhead{Med.-U} &
\colhead{Low-U} &
\colhead{Composite} &
\colhead{Observed} &
\colhead{Model/Data}\\

\colhead{} &
\colhead{} &
\colhead{Match\tablenotemark{a}} &
\colhead{Model\tablenotemark{b}} &
\colhead{Model\tablenotemark{c}} &
\colhead{Model\tablenotemark{d}} &
\colhead{Model} &
\colhead{Data\tablenotemark{e}} &
\colhead{Ratio}\\

\colhead{(1)} &
\colhead{(2)} &
\colhead{(3)} &
\colhead{(4)} &
\colhead{(5)} &
\colhead{(6)} &
\colhead{(7)} &
\colhead{(8)} &
\colhead{(9)}
}
\startdata

Ly$\alpha$  & 1216 &  1  &    25.48  &  13.99  &  11.42  &  17.81  &  14.08~$\pm$~8.35  &    1.3~$\pm$~0.8  \\
NV          & 1240 &  1  &     0.62  &   0.06  &   0.03  &   0.27  &   0.49~$\pm$~0.32  &    0.6~$\pm$~0.4  \\
CIV         & 1549 &  1  &     8.57  &   1.41  &   3.34  &   4.85  &   2.68~$\pm$~0.96  &    1.8~$\pm$~0.6  \\
CIII$]$     & 1909 &  0  &     0.76  &   1.87  &   0.15  &   0.91  &   2.45~$\pm$~1.03  &    0.4~$\pm$~0.2  \\
CII$]$      & 2326 &  1  &     0.00  &   0.05  &   1.21  &   0.38  &   0.69~$\pm$~0.26  &    0.5~$\pm$~0.2  \\
$[$NeIV$]$  & 2424 &  1  &     0.54  &   0.26  &   0.00  &   0.29  &   0.50~$\pm$~0.18  &    0.6~$\pm$~0.2  \\
MgII        & 2798 &  1  &     0.03  &   0.15  &   2.61  &   0.84  &   1.07~$\pm$~0.27  &    0.8~$\pm$~0.2  \\
$[$NeV$]$   & 3426 &  1  &     0.99  &   0.09  &   0.00  &   0.42  &   0.66~$\pm$~0.18  &    0.6~$\pm$~0.2  \\
\enddata

\tablenotetext{a}{1=Model and data line ratios relative to H$\beta$ match 
                   within a factor of 2; 0=no match.}
\tablenotetext{b}{High-U model. Boundary:  m; 
                  log$_{10}$(U)= -0.92; 
                  log$_{10}$(n$_{H}$)= 3.46~cm$^{-3}$;  
                  cloud log$_{10}$(N$_{c}$)= 21.5~cm$^{-2}$;  
                  fractional contribution = 0.40.}
\tablenotetext{c}{Medium-U model. Boundary:  m; 
                  log$_{10}$(U)= -1.92; 
                  log$_{10}$(n$_{H}$)= 4.46~cm$^{-3}$;  
                  cloud log$_{10}$(N$_{c}$)= 20.5~cm$^{-2}$;  
                  fractional contribution = 0.30.}
\tablenotetext{d}{~Low-U model. Boundary:  m; 
                  log$_{10}$(U)= -3.04; log$_{10}$(n$_{H}$)= 2.85~cm$^{-3}$;  
                  cloud log$_{10}$(N$_{c}$)= 22.0~cm$^{-2}$;  
                  fractional contribution = 0.30;
                  screen log$_{10}$(N$_{c}$)= 21.6~cm$^{-2}$.}
\tablenotetext{e}{We corrected all observed fluxes for Galactic extinction using the \citet{sav79} 
                extinction curve with E(B-V)~=~0.19.  We applied an additional correction for   
                Mrk~3 internal extinction using the LMC curve of \citet{koo81}.
                For this measurement bin the internal extinction correction 
                corresponded to E(B-V)~=~0.15.
                See Paper~I for details.}
\tablenotetext{f}{~Logarithm of model H$\beta$ flux.  Units: log$_{10}$(ergs~s$^{-1}$~cm$^{-2}$).}
\tablenotetext{g}{Observed H$\beta$ flux. Units: 10$^{-15}$~ergs~s$^{-1}$~cm$^{-2}$.}

\end{deluxetable}

\newpage

\begin{deluxetable}{lrcrrrrrr}
\label{table10}
%\rotate
\tablecolumns{9}

\tabletypesize{\scriptsize}
\tablewidth{0pc}
\tablenum{10}
\tablecaption{Line Ratios (Relative to H$\beta$) from 
Model Components, Composite and Observations for 
Position (-0$\farcs$5W,~r,~0$\farcs$2)}
\tablehead{
\colhead{} &
\colhead{} &
\colhead{Line} &
\colhead{High-U} &
\colhead{Med.-U} &
\colhead{Low-U} &
\colhead{Composite} &
\colhead{Observed} &
\colhead{Model/Data}\\

\colhead{} &
\colhead{} &
\colhead{Match\tablenotemark{a}} &
\colhead{Model\tablenotemark{b}} &
\colhead{Model\tablenotemark{c}} &
\colhead{Model\tablenotemark{d}} &
\colhead{Model} &
\colhead{Data\tablenotemark{e}} &
\colhead{Ratio}\\

\colhead{(1)} &
\colhead{(2)} &
\colhead{(3)} &
\colhead{(4)} &
\colhead{(5)} &
\colhead{(6)} &
\colhead{(7)} &
\colhead{(8)} &
\colhead{(9)}
}
\startdata

Ly$\alpha$  & 1216 &  1  &    24.30 & 24.06   &    7.18 & 19.09   &   21.30~$\pm$~2.18  &    0.9~$\pm$~0.1  \\
NV          & 1240 &  1  &     2.12 &  0.25   &    0.08 &  0.95   &    1.05~$\pm$~0.11  &    0.9~$\pm$~0.1  \\
CIV         & 1549 &  1  &    12.76 &  5.30   &    2.35 &  7.40   &    7.42~$\pm$~0.52  &    1.0~$\pm$~0.1  \\
CIII$]$     & 1909 &  1  &     0.30 &  4.06   &    0.22 &  1.41   &    1.85~$\pm$~0.13  &    0.8~$\pm$~0.1  \\
CII$]$      & 2326 &  1  &     0.00 &  0.07   &    1.41 &  0.44   &    0.57~$\pm$~0.04  &    0.8~$\pm$~0.1  \\
$[$NeIV$]$  & 2424 &  1  &     0.54 &  0.99   &    0.00 &  0.51   &    0.91~$\pm$~0.06  &    0.6~$\pm$~0.0  \\
MgII        & 2798 &  1  &     0.02 &  0.15   &    2.40 &  0.77   &    0.48~$\pm$~0.04  &    1.6~$\pm$~0.1  \\
$[$NeV$]$   & 3426 &  1  &     1.64 &  0.39   &    0.00 &  0.77   &    1.04~$\pm$~0.05  &    0.7~$\pm$~0.0  \\
\enddata

\tablenotetext{a}{1=Model and data line ratios relative to H$\beta$ match 
                   within a factor of 2; 0=no match.}
\tablenotetext{b}{High-U model. Boundary:  m; 
                  log$_{10}$(U)= -0.38; 
                  log$_{10}$(n$_{H}$)= 2.86~cm$^{-3}$;  
                  cloud log$_{10}$(N$_{c}$)= 22.0~cm$^{-2}$;  
                  fractional contribution = 0.40.}
\tablenotetext{c}{Medium-U model. Boundary:  m; 
                  log$_{10}$(U)= -1.88; 
                  log$_{10}$(n$_{H}$)= 4.36~cm$^{-3}$;  
                  cloud log$_{10}$(N$_{c}$)= 20.0~cm$^{-2}$;  
                  fractional contribution = 0.30.}
\tablenotetext{d}{~Low-U model. Boundary:  m; 
                  log$_{10}$(U)= -2.78; log$_{10}$(n$_{H}$)= 2.50~cm$^{-3}$;  
                  cloud log$_{10}$(N$_{c}$)= 22.5~cm$^{-2}$;  
                  fractional contribution = 0.30;
                  screen log$_{10}$(N$_{c}$)= 21.6~cm$^{-2}$.}
\tablenotetext{e}{We corrected all observed fluxes for Galactic extinction using the \citet{sav79} 
                extinction curve with E(B-V)~=~0.19.  We applied an additional correction for   
                Mrk~3 internal extinction using the LMC curve of \citet{koo81}.
                For this measurement bin the internal extinction correction 
                corresponded to E(B-V)~=~0.16.
                See Paper~I for details.}
\tablenotetext{f}{~Logarithm of model H$\beta$ flux.  Units: log$_{10}$(ergs~s$^{-1}$~cm$^{-2}$).}
\tablenotetext{g}{Observed H$\beta$ flux. Units: 10$^{-15}$~ergs~s$^{-1}$~cm$^{-2}$.}

\end{deluxetable}

\newpage

\begin{deluxetable}{lrcrrrrrr}
\label{table11}
%\rotate
\tablecolumns{9}

\tabletypesize{\scriptsize}
\tablewidth{0pc}
\tablenum{11}
\tablecaption{Line Ratios (Relative to H$\beta$) from 
Model Components, Composite and Observations for 
Position (-0$\farcs$7W,~r,~0$\farcs$3)}
\tablehead{
\colhead{} &
\colhead{} &
\colhead{Line} &
\colhead{High-U} &
\colhead{Med.-U} &
\colhead{Low-U} &
\colhead{Composite} &
\colhead{Observed} &
\colhead{Model/Data}\\

\colhead{} &
\colhead{} &
\colhead{Match\tablenotemark{a}} &
\colhead{Model\tablenotemark{b}} &
\colhead{Model\tablenotemark{c}} &
\colhead{Model\tablenotemark{d}} &
\colhead{Model} &
\colhead{Data\tablenotemark{e}} &
\colhead{Ratio}\\

\colhead{(1)} &
\colhead{(2)} &
\colhead{(3)} &
\colhead{(4)} &
\colhead{(5)} &
\colhead{(6)} &
\colhead{(7)} &
\colhead{(8)} &
\colhead{(9)}
}
\startdata

Ly$\alpha$  & 1216 &   1  &  24.38    &   5.26  &  11.76  & 11.36   &  20.81~$\pm$~1.48  &    0.5~$\pm$~0.0  \\
NV          & 1240 &   1  &   3.27    &   0.16  &   0.07  &  0.75   &   0.98~$\pm$~0.42  &    0.8~$\pm$~0.3  \\
CIV         & 1549 &   1  &  20.71    &   2.80  &   3.54  &  6.64   &   5.04~$\pm$~0.30  &    1.3~$\pm$~0.1  \\
CIII$]$     & 1909 &   0  &   0.55    &   1.77  &   0.22  &  0.98   &   2.09~$\pm$~0.14  &    0.5~$\pm$~0.0  \\
CII$]$      & 2326 &   1  &   0.00    &   0.02  &   1.46  &  0.52   &   0.62~$\pm$~0.03  &    0.8~$\pm$~0.0  \\
$[$NeIV$]$  & 2424 &   1  &   0.94    &   0.57  &   0.00  &  0.45   &   0.78~$\pm$~0.08  &    0.6~$\pm$~0.1  \\
MgII        & 2798 &   1  &   0.04    &   0.09  &   2.69  &  0.99   &   0.90~$\pm$~0.04  &    1.1~$\pm$~0.0  \\
$[$NeV$]$   & 3426 &   1  &   2.80    &   0.43  &   0.00  &  0.75   &   0.76~$\pm$~0.03  &    1.0~$\pm$~0.0  \\
\enddata

\tablenotetext{a}{1=Model and data line ratios relative to H$\beta$ match 
                   within a factor of 2; 0=no match.}
\tablenotetext{b}{High-U model. Boundary:  m; 
                  log$_{10}$(U)= -0.45; 
                  log$_{10}$(n$_{H}$)= 2.57~cm$^{-3}$;  
                  cloud log$_{10}$(N$_{c}$)= 21.8~cm$^{-2}$;  
                  fractional contribution = 0.20.}
\tablenotetext{c}{Medium-U model. Boundary:  m; 
                  log$_{10}$(U)= -1.45; 
                  log$_{10}$(n$_{H}$)= 3.57~cm$^{-3}$;  
                  cloud log$_{10}$(N$_{c}$)= 21.0~cm$^{-2}$;  
                  fractional contribution = 0.45.}
\tablenotetext{d}{~Low-U model. Boundary:  m; 
                  log$_{10}$(U)= -2.90; log$_{10}$(n$_{H}$)= 2.25~cm$^{-3}$;  
                  cloud log$_{10}$(N$_{c}$)= 22.0~cm$^{-2}$;  
                  fractional contribution = 0.35;
                  screen log$_{10}$(N$_{c}$)= 21.6~cm$^{-2}$.}
\tablenotetext{e}{We corrected all observed fluxes for Galactic extinction using the \citet{sav79} 
                extinction curve with E(B-V)~=~0.19.  We applied an additional correction for   
                Mrk~3 internal extinction using the LMC curve of \citet{koo81}.
                For this measurement bin the internal extinction correction 
                corresponded to E(B-V)~=~0.14.
                See Paper~I for details.}
\tablenotetext{f}{~Logarithm of model H$\beta$ flux.  Units: log$_{10}$(ergs~s$^{-1}$~cm$^{-2}$).}
\tablenotetext{g}{Observed H$\beta$ flux. Units: 10$^{-15}$~ergs~s$^{-1}$~cm$^{-2}$.}

\end{deluxetable}

\newpage

\begin{deluxetable}{lrcrrrrrr}
\label{table12}
%\rotate
\tablecolumns{9}

\tabletypesize{\scriptsize}
\tablewidth{0pc}
\tablenum{12}
\tablecaption{Line Ratios (Relative to H$\beta$) from 
Model Components, Composite and Observations for 
Position (-1$\farcs$0W,~r,~0$\farcs$3)}
\tablehead{
\colhead{} &
\colhead{} &
\colhead{Line} &
\colhead{High-U} &
\colhead{Med.-U} &
\colhead{Low-U} &
\colhead{Composite} &
\colhead{Observed} &
\colhead{Model/Data}\\

\colhead{} &
\colhead{} &
\colhead{Match\tablenotemark{a}} &
\colhead{Model\tablenotemark{b}} &
\colhead{Model\tablenotemark{c}} &
\colhead{Model\tablenotemark{d}} &
\colhead{Model} &
\colhead{Data\tablenotemark{e}} &
\colhead{Ratio}\\

\colhead{(1)} &
\colhead{(2)} &
\colhead{(3)} &
\colhead{(4)} &
\colhead{(5)} &
\colhead{(6)} &
\colhead{(7)} &
\colhead{(8)} &
\colhead{(9)}
}
\startdata

Ly$\alpha$  & 1216 &  0  &  32.25   &   12.53 &   24.86 &  19.43  &   42.54~$\pm$~13.18  &   0.5~$\pm$~0.2  \\
NV          & 1240 &  1  &   7.59   &    0.18 &    0.08 &   0.88  &    0.86~$\pm$~0.28  &    1.0~$\pm$~0.3  \\
CIV         & 1549 &  1  &  14.05   &    3.63 &    6.08 &   5.65  &    5.32~$\pm$~1.00  &    1.1~$\pm$~0.2  \\
CIII$]$     & 1909 &  0  &   0.09   &    2.57 &    0.30 &   1.41  &    4.74~$\pm$~6.71  &    0.3~$\pm$~0.4  \\
CII$]$      & 2326 &  0  &   0.00   &    0.03 &    1.53 &   0.63  &    1.66~$\pm$~2.11  &    0.4~$\pm$~0.5  \\
$[$NeIV$]$  & 2424 &  1  &   0.17   &    0.77 &    0.00 &   0.40  &    0.76~$\pm$~0.19  &    0.5~$\pm$~0.1  \\
MgII        & 2798 &  1  &   0.00   &    0.10 &    2.80 &   1.17  &    1.91~$\pm$~0.24  &    0.6~$\pm$~0.1  \\
$[$NeV$]$   & 3426 &  1  &   2.89   &    0.41 &    0.00 &   0.49  &    0.53~$\pm$~0.08  &    0.9~$\pm$~0.1  \\
\enddata

\tablenotetext{a}{1=Model and data line ratios relative to H$\beta$ match 
                   within a factor of 2; 0=no match.}
\tablenotetext{b}{High-U model. Boundary:  m; 
                  log$_{10}$(U)= -0.24; 
                  log$_{10}$(n$_{H}$)= 2.08~cm$^{-3}$;  
                  cloud log$_{10}$(N$_{c}$)= 21.0~cm$^{-2}$;  
                  fractional contribution = 0.10.}
\tablenotetext{c}{Medium-U model. Boundary:  m; 
                  log$_{10}$(U)= -1.66; 
                  log$_{10}$(n$_{H}$)= 3.50~cm$^{-3}$;  
                  cloud log$_{10}$(N$_{c}$)= 20.5~cm$^{-2}$;  
                  fractional contribution = 0.50.}
\tablenotetext{d}{~Low-U model. Boundary:  m; 
                  log$_{10}$(U)= -2.93; log$_{10}$(n$_{H}$)= 2.00~cm$^{-3}$;  
                  cloud log$_{10}$(N$_{c}$)= 21.5~cm$^{-2}$;  
                  fractional contribution = 0.40;
                  screen log$_{10}$(N$_{c}$)= 21.6~cm$^{-2}$.}
\tablenotetext{e}{We corrected all observed fluxes for Galactic extinction using the \citet{sav79} 
                extinction curve with E(B-V)~=~0.19.  We applied an additional correction for   
                Mrk~3 internal extinction using the LMC curve of \citet{koo81}.
                For this measurement bin the internal extinction correction 
                corresponded to E(B-V)~=~0.12.
                See Paper~I for details.}
\tablenotetext{f}{~Logarithm of model H$\beta$ flux.  Units: log$_{10}$(ergs~s$^{-1}$~cm$^{-2}$).}
\tablenotetext{g}{Observed H$\beta$ flux. Units: 10$^{-15}$~ergs~s$^{-1}$~cm$^{-2}$.}

\end{deluxetable}

\newpage

%%%%%%%%%%%%%%%%%%%%%%%%%%%%%%%%%%%%%%%%%%%%%%%%%%%%%%%%%%%%%%%%%%%%%%%%%%%%%%

%write_cloudy_tex_table.pro
\newpage
\begin{deluxetable}{llcclrcrrrrr}
\label{table13}
\tabletypesize{\scriptsize}
\rotate
%\tablewidth{560pt}
\tablewidth{0pc}
\tablenum{13}
\tablecolumns{12}
\tablecaption{CLOUDY Photoionization Model Output}
\tablehead{
\colhead{Bin} &
\colhead{Model} &
\colhead{Bound}\tablenotemark{a} &
\colhead{U} &
\colhead{H$\beta$} &
\colhead{Emitting} &
\colhead{Component} & 
\colhead{Bin} &
\colhead{Component} &
\colhead{Bin} &
\colhead{Filling} &
\colhead{T$_{avg.}$} \\
\colhead{} &
\colhead{} &
\colhead{} &
\colhead{} &
\colhead{Flux} &
\colhead{Area} &
\colhead{Depth} & 
\colhead{Depth} &
\colhead{Height} & 
\colhead{Height} &
\colhead{Factor} &
\colhead{} \\
\colhead{} &
\colhead{} &
\colhead{} &
\colhead{} &
\colhead{(erg s$^{-1}$ cm$^{-2}$)} &
\colhead{(pc$^{2}$)} &
\colhead{(pc)} &
\colhead{(pc)} &
\colhead{(pc)} &
\colhead{(pc)} &
\colhead{} &
\colhead{(K)} \\
\colhead{(1)} &
\colhead{(2)} &
\colhead{(3)} &
\colhead{(4)} &
\colhead{(5)} &
\colhead{(6)} &
\colhead{(7)} &
\colhead{(8)} &
\colhead{(9)} &
\colhead{(10)} &
\colhead{(11)} &
\colhead{(12)} 
}
\startdata
( 0$\farcs$5E, r, 0$\farcs$3) & high & m & 10$^{-0.29}$ & 2.23 $\times$ 10$^{+00}$ &     211 &       8 &  26 &     19 &  79 & 10$^{-1.12}$ & 17000 \\
                             & med. & m & 10$^{-2.23}$ & 6.06 $\times$ 10$^{-01}$ &     777 &      29 &  26 &   $<$1 &  79 & 10$^{-5.00}$ & 12000 \\
                             & low  & m & 10$^{-2.69}$ & 3.63 $\times$ 10$^{-01}$ &    1111 &      42 &  42 &     15 &  79 & 10$^{-0.71}$ &  9000 \\
( 0$\farcs$3E, b, 0$\farcs$3) & high & m & 10$^{-0.45}$ & 3.33 $\times$ 10$^{+00}$ &      44 &       1 &  14 &      1 &  66 & 10$^{-2.70}$ & 16000 \\
                             & med. & i & 10$^{-1.95}$ & 1.04 $\times$ 10$^{+01}$ &      25 &       0 &  14 &   $<$1 &  66 & 10$^{-5.45}$ & 10000 \\
                             & low  & m & 10$^{-3.09}$ & 2.28 $\times$ 10$^{-01}$ &     777 &      29 &  25 &      2 &  66 & 10$^{-1.38}$ &  8000 \\
( 0$\farcs$3E, r, 0$\farcs$3) & high & m & 10$^{-0.89}$ & 3.43 $\times$ 10$^{+00}$ &      74 &       2 &  12 &   $<$1 &  66 & 10$^{-3.42}$ & 15000 \\
                             & med. & i & 10$^{-2.39}$ & 1.43 $\times$ 10$^{+01}$ &      32 &       1 &  12 &   $<$1 &  66 & 10$^{-6.09}$ &  9000 \\
\enddata
\tablenotetext{a}{``m'' indicates that the model cloud is matter bounded and ``i'' indicates that it is 
ionization bounded.}
%\tablenotetext{b}{}
\end{deluxetable}

%%%%%%%%%%%%%%%%%%%%%%%%%%%%%%%%%%

\clearpage
\begin{figure}
\plotone{f1.eps}
\\Fig.~1.
\end{figure}

\clearpage
\begin{figure}
\plotone{f2.eps}
\\Fig.~2.
\end{figure}

\clearpage
\begin{figure}
\plotone{f3.eps}
\\Fig.~3.
\end{figure}

\clearpage
\begin{figure}
\plotone{f4.eps}
\\Fig.~4.
\end{figure}

\clearpage
\begin{figure}
\plotone{f5.eps}
\\Fig.~5.
\end{figure}

\clearpage
\begin{figure}
\plotone{f6.eps}
\\Fig.~6.
\end{figure}

\clearpage
\begin{figure}
\plotone{f7.eps}
\\Fig.~7.
\end{figure}


\begin{thebibliography}{}


\bibitem[Adams(1977)]{ada77} Adams, T.F. 1977, \apjs, 33, 19

\bibitem[Alexander et al.(1999)]{ale99} Alexander, T., Sturm, E., Lutz, D.,
Sternberg, A., Netzer, H., \& Genzel, R. 1999, \apj, 512, 204

\bibitem[Antonucci(1993)]{ant93} Antonucci, R.R.J. 1993, \araa, 31,
473

\bibitem[Barvainis(1987)]{bar87} Barvainis, R. E. 1987, \apj, 320, 537

\bibitem[Binette \& Robinson(1987)]{bin87} Binette, L., \& Robinson, A. 
1987, \aap, 177, 11

\bibitem[Binette, Wilson \& Storchi-Bergmann(1996)]{bin96} Binnette, L., 
Wilson, A. S. \& Storchi-Bergmann, T. 1996, \aap, 312, 365

\bibitem[Binney \& Merrifield(1998)]{bin98} 
Binney, J. \& Merrifield, M. 1998 in ``Galactic Astronomy'' 
(Princeton University Press)

\bibitem[Collins et al.(2005)]{col05} Collins, N.R., Kraemer, S.B., 
Crenshaw, D.M., Ruiz, J., Deo, R. \& Bruhweiler, F.C, \apj, 619, 116 (Paper I)

\bibitem[Crenshaw et al.(2003)]{cre03} Crenshaw, D.M. et al. 2003,
\apj, 594, 116

\bibitem[Evans et al.(1993)]{eva93} Evans, I.N., Tsvetanov, Z., Kriss,
G.A., Ford, H.C., Caganoff, S. \& Koratkar, A.P. 1993, \apj, 417, 82

\bibitem[Feldmeier et al.(1999)]{fel99} Feldmeier, J.J., 
Brandt, W.N., Elvis, E., Fabian, A.C., 
Iwasawa, K. \& Mathur, S. 1999, \apj, 510, 167

\bibitem[Ferland \& Netzer(1983)]{fer83} Ferland, G.J. \& Netzer,
H. 1983, \apj, 264, 105

\bibitem[Ferland et al.(1998)]{fer98} Ferland, G. J.,  Korista, K.T., 
Verner, D.A., Ferguson, J.W., Kingdon, J.B. \& Verner,  E.M. 1998, 
\pasp, 110, 761

\bibitem[Grevesse \& Anders(1989)]{gre89} Grevasse, N. \& 
Anders, E. 1989, ``Cosmic Abundances of Matter'', AIP Conf. Proc. 183,
ed. C.J. Waddington (New York: AIP)

\bibitem[Konigl \& Kartje(1994)]{kon94} Konigl, A. \& Kartje, J.F. 
1994, \apj, 434 446

\bibitem[Koornneef \& Code(1981)]{koo81} Koornneef, J. \& Code,
A.D. 1981, \apj, 247, 860

\bibitem[Kraemer \& Harrington(1986)]{kra86} Kraemer,S.B. \&
Harrington, J.P. 1986, \apj, 307, 478

\bibitem[Kraemer et al.(2000)]{kra00} Kraemer, S.B., Crenshaw, D.M.,
Hutchings, J.B., Gull, T.R., Kaiser, M.E., Nelson, C.H. \& Weistrop,
D. 2000, \apj, 531, 278

\bibitem[Kraemer \& Crenshaw(2000a)]{kc00a} Kraemer, S.B. \& Crenshaw,
D.M. 2000, \apj, 532, 256

\bibitem[Kraemer \& Crenshaw(2000b)]{kc00b} Kraemer, S.B. \& Crenshaw,
D.M. 2000, \apj, 544, 763

\bibitem[Kraemer et al.(2001)]{kra01} Kraemer, S.B. et al.  
2001, \apj, 551, 671

\bibitem[Kraemer et al.(2004)]{kra04} Kraemer, S.B.,
George, I.M., Crenshaw, D.M., \& Gabel, J.R.  2001, \apj, 607, 794

\bibitem[Kraemer, Schmitt, \& Crenshaw(2008)]{kra08} Kraemer, S.B., 
Schmitt, H.R., \& Crenshaw, D.M.  2008, \apj, 679, 1128

\bibitem[Mathis, Rumpl, \& Nordsieck(1977)]{mat77} Mathis, J.S., Rumpl, W., \&
Nordsieck, K.H 1977, \apj, 217, 425

\bibitem[Matt et al.(2000)]{mat00} Matt, G., Fabian, A.C., Guainazzi, M., 
Iwasawa, K., Bassani, L. \& Malaguti, G. 2000, \mnras, 318, 173

\bibitem[M\'{e}lendez et al.(2008)]{mel08} M\'{e}lendez, M., Kraemer, S. B., 
Armentrout, B. K., Deo, R. P., Crenshaw, D. M., Schmitt, H. R., 
Mushotzky, R. F., Tueller, J., Markwardt, C. B., Winter, L.  
2008, \apj, 682, 94

\bibitem[Neugebauer et al.(1976)]{neu76} Neugebauer, G., Becklin, J.B., 
Oke, J.B \& Searle, L. 1976, \apj, 205, 29

\bibitem[Noordermeer et al. (2005)]{noo05} Norrdermeer, E., 
van der Hulst, J.M., Sancisi, R., Swaters, R.A. \& van Albada, T.S. 
2005, \aap, 442, 137

\bibitem[Osterbrock(1989)]{ost89} Osterbrock, D.E. 1989, Astrophysics
of Gaseous Nebulae and Active Galactic Nuclei (Mill Valley: University Science
Books)

\bibitem[Peterson(1997)]{pet97} Peterson, B.M. 1997, An Introduction to 
Active Galactic Nuclei (Cambridge: Cambridge University Press)

\bibitem[Peterson et al.(2004)]{pet04} Peterson, B.M., 
Ferrarese, L., Gilbert, K.M., Kaspi, S., Malkan, M.A., Maoz, D., 
Merritt, D., Netzer, H., Onken, C.A., Pogge, R.W., Vestergaard, M. 
\& Wandel, A. 2004, \apj, 613, 682 

\bibitem[Rieke(1978)]{rie78} Rieke, G.H. 1978, \apj, 226, 550

\bibitem[Ruiz et al.(2001)]{rui01} Ruiz, J.R., Crenshaw, D.M.,
Kraemer, S.B., Bower, G.A., Gull, T.R., Hutchings, J.B., Kaiser,
M.E. \& Weistrop, D. 2001, \aj, 122, 2961

\bibitem[Sako et al.(2000)]{sak00} Sako, M., Kahn, S.M., Paerels,
F. \& Liedahl, A. 2000, \apjl, 543, 115

\bibitem[Savage \& Mathis(1979)]{sav79} Savage, B.D. \& Mathis, J.S
1979, \araa, 17, 73

\bibitem[Schlegel et al.(1998)]{sch98} Schlegel, D.J., Finkbeiner,
D.P. \& Davis, M. 1998, \apj, 500, 525

\bibitem[Schmidt \& Miller(1985)]{sch85} Schmidt, G.D. \& Miller,
J.S. 1985, \apj, 290, 517

\bibitem[Schmitt \& Kinney(1996)]{sch96} Schmitt, H.R., \& Kinney,
A.L.  1996, \apj, 463, 498

\bibitem[Schmitt \& Kinney(2000)]{sch00} Schmitt, H.R., \& Kinney,
A.L. 2000, \apjs, 128, 479

\bibitem[Schmitt et al.(2003)]{sch03} Schmitt, H.R., Donley, J.L., 
Antonucci, R.R.J., Hutchings, J.B., \& Kinney, A.L. 2003, \apjs, 148, 327

\bibitem[Seab \& Shull(1983)]{sea83} Seab, C.G. \& Shull, J.M. 1983,
\apj, 275, 652

\bibitem[Seaton(1978)]{sea78} Seaton, M.J. 1978, \mnras, 185, 5P

\bibitem[Snow \& Witt(1996)]{sno96} Snow, T.P. \& Witt, A.N. 1996,
\apj, 65L

\bibitem[Staveley-Smith et al.(2003)]{sta03} Staveley-Smith, L., 
Kim, S., Calbretta, M.R., Haynes, R.F. \& Kesteven, M.J. 
2003, \mnras, 339, 87

\bibitem[Swaters et al.(2002)]{swa02} Swaters, R.A., van Albada, T.S., 
van der Hulst, J.M. \& Sancisi, R. 2002, \aap, 390, 829

\bibitem[Tifft \& Cocke(1988)]{tif88} Tifft, W.G. \& Cocke, W.J. 1988,
\apjs, 67, 1

\bibitem[Turner et al. (1997)]{tur97} Turner, J., George, I.M., Nandra, K. 
\& Mushotzky, R.F. 1997, \apj, 488, 164

\bibitem[Ulvestad et al.(1998)]{ulv98} Ulvestad, J.S., 
Roy, A.L., Colbert, E.J.M., Wilson, A.S, 1998, \apj, 496, 196

\bibitem[Weedman et al.(2005)]{wee05} Weedman, D.W, Hao, L., Higdon,
S.J.U., Devost, D., Yanling, W., Charmandaris, V., Brandl, B., Bass,
E. \& Houck, J.R.  2005, \apj, 633, 706

\bibitem[Welch et al.(1998)]{wel98} Welch, G.A., Sage, L.J. \& 
Mitchell, G.F. 1998, \apj, 499, 209

\bibitem[Woo \& Urry(2002)]{woo02} Woo, J.-H., \& Urry, C.M. 2002, \apj, 579,
530


\end{thebibliography}
\end{document}